\documentclass[12pt,aps,prd,superscriptaddress,showpacs,amsmath,amssymb]{revtex4}
\pdfoutput=1
\usepackage{epsf}
\usepackage{latexsym}
\usepackage{amsmath}
\usepackage{amsfonts}
\usepackage{amssymb}
\usepackage{graphicx}
\def\:{\ \ \ }
\newcommand{\be}{\begin{eqnarray}}
\newcommand{\ee}{\end{eqnarray}}

\usepackage{lscape}
\usepackage{graphics}
\usepackage{epsfig}
\usepackage{graphicx}
\usepackage{rotate}
\usepackage{color}
\usepackage{amsmath, amsthm} 

\def\ARAA{{\it Annual Rev. of Astron. \& Astrophys.} }
\def\ApJ{{\it Astrophys. J.} }
\def\ApJL{{\it Astrophys. J. Letters} }
\def\ApJS{{\it Astrophys. J. Suppl.} }

\def\AA{{\it Astron. \& Astroph.} }

\def\MNRAS{{\it Month. Not. Roy. Astr. Soc.} }
\def\Nature{{\it Nature} }
\def\NewAR{{\it New Astron. Rev.} }

\def\PLB{{\it Phys. Lett.}{\,}{\bf B} }
\def\PR{{\it Phys. Rev.} }
\def\PRD{{\it Phys. Rev.} {\bf D} }
\def\PRL{{\it Phys. Rev. Letters} }
\def\RMP{{\it Rev. Mod. Phys.} }
\def\Science{{\it Science}Ê}

\def\ha{\hangindent=1.15in}
\def\simle{\lower 2pt \hbox {$\buildrel < \over {\scriptstyle \sim }$}}
\def\simge{\lower 2pt \hbox {$\buildrel > \over {\scriptstyle \sim }$}}

\begin{document}
\title{ A graviton statistics approach to dark energy, black holes and inflation}
\author{Peter L. Biermann}
\email{plbiermann@mpifr-bonn.mpg.de}
\affiliation{Department of Physics and Astronomy, The University
of Alabama, Box 870324, Tuscaloosa, AL 35487-0324, USA}
\affiliation{MPI for Radioastronomy, Bonn, Germany}
\affiliation{Karlsruhe Institute of Technology (KIT) - Institut f{\"u}r Kernphysik, Germany}
\affiliation{Department of Physics, University of Alabama at Huntsville, AL, USA}
\affiliation{Department of Physics \& Astronomy, University of Bonn, Germany}
\author{Benjamin C. Harms}
\email{bharms@bama.ua.edu}
\affiliation{Department of Physics and Astronomy, The University
of Alabama, Box 870324, Tuscaloosa, AL 35487-0324, USA}
%
%
%

\begin{abstract}

We derive two new equations of quantum gravity and combine them with reinterpretations of previously proposed concepts of dark energy, black holes, inflation, the arrow of time and the characteristic energy at which rest-mass first manifests itself into a theory which may be a first step toward a comprehensive description of all these phenomena.  The resulting theory also predicts new tests which can be experimentally checked within just a few years. The two new equations are :  A)  a creation equation to give stimulated emission for any surface filled with gravitons, pulling energy from a background, and B) the association of an outgoing soliton wave of gravitons, a ``shell front''  with a large Lorentz factor derived from the uncertainties in both space and time. We model the background as a strong gravity brane, a Planck length apart from our brane in a fifth dimension.  These new equations are combined with the common notions of an all-pervasive background of gravitons at the Planck limit, the ``Planck sea''; the identification of the thermodynamic limit with the emission of gravitons in a ``shell front'', i.e. what is usually called the entropy of black holes is identified with the outgoing gravitons; the concept of black holes as a membrane full of gravitons at a large Lorentz factor, the ``Planck shell'';  the emission of gravitons created in a "horizon shell" during inflation, which lose energy adiabatically with the Planck time.  These equations result in stimulated emission of gravitons by interaction with the background, the ``Planck sea'', to describe dark energy, black holes, and the inflationary period of the universe.   These proposals lead to gravitational waves constituting dark energy.  These waves should be detectable within a few years with pulsar timing arrays. The extremely high, but finite Lorentz factor for signal propagation may be expected to have further consequences in particle interactions.
\end{abstract}
\maketitle

\section{Introduction}

The goal of this paper is to provide the first step towards a description of a comprehensive model of the birth and evolution of the universe.  This model describes how the interaction with an all-pervasive background field drives inflation, black holes, dark energy, and the arrow of time.  The background field is assumed to be the always the same one, sending energy into our world by stimulated emission.  For a review of the standard quantum mechanical picture of the evolution of the universe, see \cite{gri}.  

One of the new equations is the creation equation derived in section III. The other is the Lorentz factor expression for the motion of the shell front of emitted gravitons derived in section IV. The form of the creation equation is similar to the one obtained from the non-relativistic form of the Boltzmann equation (the Kompaneets equation), and the square of the phase space density occurs as  in both the relativistic and non-relativistic forms of the Boltzmann equation.  The argument for the Lorentz factor is similar to the connection between the scale of the universe with the Planck scales as done in \cite{gibhaw77,gibhaw79}, as well as the equation describing the time delay of arrival between a photon and an ultra high energy particle emitted simultaneously from across the universe.  Our concept is quite specific, and leads to predictions, which can be either verified or disproved within a few years, probably within existing experimental plans.  We generally use a micro-canonical ensemble approach, a statistics method (e.g. \cite{har92,har93}), instead of thermodynamics.  It is well known that gravitational systems are never in full thermodynamic equilibrium, and so for gravitational systems the expression giving no change of entropy
\begin{equation}
d S \; = \; 0
\end{equation}
never holds.  The microcanonical ensemble is thus the appropriate ensemble to use for the description of gravitational systems.

The creation equation (\ref{creationeq}) is  a combination of the ``holographic principle" \cite{bek7, bou} and the relativistic form of the Kompaneets equation, \cite{kom,ryb,zel}. 
In our model a quasi static background field exists which can emit gravitons via stimulated emission into our universe.  The connection between the background field and our world is in ``shells" whose thicknesses equal the Planck scale $ \sim \, 10^{-33} \, {\rm cm}$ in the comoving frame.  The thickness of a shell is determined by the size of the fifth dimension.  This background field and the emitted gravitons are taken to be real. The background field will be referred to as the ``Planck sea", and it is described in the observer frame.  Its energy spectrum is a Planck spectrum at maximal energy density, peaking at the Planck energy as seen by an observer in our world.   This corresponds rather closely to the notion of an ``all-pervading background field", used quite often in the literature ( e.g. \cite{fri}).  The distinction between our model and what is described in \cite{fri} is that we assume that the background is important at all times.  We will show below how the source of energy in our universe can be incorporated into the energy-momentum tensor and the Einstein equations, and the form is well known.

The approach described here may also explain the inflationary period, with new gravitons produced in a horizon shell, provided that the newly created gravitons lose their energy adiabatically with a time scale proportional to the Planck time, independent of redshift within the inflationary period; this is just a specific application of standard models. In this concept the universe starts with one Planck energy graviton in one Planck volume; this is conceptually similar to some ideas of Lema\^{i}tre \cite{lem}.  In our model the inflationary period ends when the number density distribution of the created gravitons begins to deviate significantly from the Bose-Einstein distribution, which coincides with the first appearance of the Higgs boson.

The gravitational waves constituting dark energy in our concept may be measurable by pulsar timing arrays.  The continuous creation of gravitons due to the interaction of Planck shells with the background implies an energy density - pressure relation 
\be 
P_{DE} = -\rho_{DE}\,c^2 \, .
\ee
The gravitational waves produced in the formation of the first generation of stellar black holes should be measurable by gravitational wave interferometers.  The soliton-like passage of a single graviton shell front may be measurable by Lunar laser ranging, as well as by ultra-precise timing measurements.

This paper is organized as follows:  In section II we give a detailed description of the background Planck sea.  In section III we derive the creation equation.  In section IV we discuss dark energy, in section V black holes, in section VI inflation, in section VII predictions for possible observations, and in section VIII we give some further consequences such as an emergent arrow of time, and the first appearance of rest mass, as well as summarize.  For numerical illustration we use black holes at $3 \, 10^{6} \, M_{\odot}$ and a formation redshift of 50.

\section{The Planck sea}

We interpret the usual notion of a quasi static background field as one filled with gravitons.  Models with a static background have been treated in the literature, see for example \cite{Binetruy00}.  Strictly speaking 
the background is evolving on a time scale which is slower by the energy 
density ratio, $\epsilon_b\,=\,10^{115.3}$ erg/cm$^3$ in the background (the Planck 
limit) to $\epsilon_{DE}\,=\,10^{-8}$ erg/cm$^3$ in our universe (the dark energy) as we 
will show below in section IV.C and VI.  A possible model which has the features we require is two four-branes embedded in a five-dimensional bulk and separated by the Planck length; such a model is inspired by but different from the Randall-Sundrum models \cite{RS99a,RS99b}.  A metric which describes the transfer of energy from a four-brane where gravity is strong to the brane on which we live is given by
\be
ds^2 = -e^{(u/l)^m\,t/\psi}\,c^2\,dt^2\,+\,{{\rm e}^{ \left( 1-b \left( {\frac {u}{l}} \right) ^{n} \right)\,2\,t/ {
\alpha}{{\it \tau_H}}}}\,{{\rm e}^{- \left( 
{\frac {u}{l}} \right) ^{k} \left( 1-{\frac {t}{\phi}} \right) }}\,dx_i\,dx^i + {{\rm e}^{ \left( {\frac {u}{l}}
 \right) ^{p}\,t/{\beta}}}\,du^2 \,
\label{metric} 
 \ee
where $i\,=\,1,2,3$, $\tau_H$ is the Hubble time, $l$ is the Planck length, $u$ is the coordinate in the fifth dimension, and the remaining, non-coordinate quantities are arbitrary parameters.  This metric is not an exact solution of Einstein's equations since the five-dimensional covariant divergence of the energy-momentum tensor does not vanish everywhere.  However, the five-dimensional divergence vanishes on the weak-gravity brane ($u = 0$), and it is approximately zero on the strong-gravity brane ($u = l$) for large $\beta/\tau_H$ and $\psi/\tau_H$.  This metric describes a weak-gravity brane which is expanding with time and a strong-gravity brane which is contracting with time, albeit very slowly for the latter brane.  The cosmological constant measured on the weak-gravity brane is
\be 
\Lambda_{weak-gravity}\,=\,-3\,{\frac {1}{{\alpha}^{2}{{\it \tau_H}}^{2}}}
\ee

The evolution of the strong-gravity brane is determined  by the relative sizes of the parameters $\alpha\, , \tau_H$ and $\beta$.  The metric in  Eq.(\ref{metric}) bears a resemblance to the first Randall - Sundrum model except that the metric tensor elements in Eq.(\ref{metric}) are time-dependent, and the tensor element $g_{44}$ depends exponentially on the 5th coordinate, $u$ \cite{RS99a,RS99b}.  This metric is clearly an approximate description of the current geometry of the universe and does not describe the geometry of the universe for all epochs.  The vanishing of terms linear and quadratic in time appearing in the Einstein tensor, $G_{00}$, evaluated on the strong-gravity brane requires that
\be 
\phi\,=\,\frac{k\,\alpha\,\tau_H}{2\,b\,n}\, , \hspace*{5mm}{\rm or} \hspace*{5mm} \phi\,=\,\frac{2\,k\,\alpha\,\tau_H\,\beta}{4\,\beta\,n\,b\,+\,p\,\alpha\,\tau_H}\, ,
\ee
and, using the first expression for $\phi$ above,
\be 
k\,=\,0\, , \hspace*{5mm}{\rm or} \hspace*{5mm} k\,=\,\frac{4\,b\,n^2\,\beta}{4\,\beta\,n\,b\,+\,p\,\alpha\,\tau_H} \, .
\ee

 The scale for changes in time is given by the ratio
\be 
\tau_{change}\,=\,\frac{\epsilon_b}{\epsilon_{DE}}\,\tau_H
\ee
The scale of changes in time is thus, including factors of $(4\,\pi)^3$, 
\be 
\tau_{change}\,\approx\,10^{120}\,\tau_H\, .
\ee
Factors of this magnitude recur often in 
cosmology, as already noted by Dirac \cite{dirac}, who showed that the mass of 
the universe is about $10^{80}$ the mass of a proton, while other 
factors such as the ratio of the electric to the gravitational force are on the order of 
$10^{40}$.  The factor here is approximately $10^{120}$ \cite{zeldovich,weinberg}, and the long-sought explanation for this large factor emerges naturally from our 
model. We call this field the ``Planck sea", and assume that the gravitons in this sea have a Planck spectrum at maximal energy density peaked at the Planck energy.  The Planck sea is in equilibrium with respect to gravitational collapse, because the  free-fall time scale and pressure wave time scale are equal at the Planck length. Larger scales are not  causally connected.  As noted in the Introduction, the background field and the emitted gravitons are not virtual but are taken to be real.  The background field spectrum is modified at the (lower) energy where the coupling strength of the four natural forces diverge, about a factor of order $10^{18}$ from the maximum in the very early universe, as suggested by observations of the end of the inflationary period.  This field provides the energy for the gravitons created by stimulated emission in our universe, as we discuss below.  The total of the background field energy and the energy in our observed universe is conserved.

We assume a specific field with a spectrum of
\be
\frac{1}{\pi^2} \, \frac{\epsilon^{3}}{e^\epsilon - 1} \, d \epsilon
\label{Planckseasp}
\ee
in the observer frame where 
\be
\epsilon \; = \; \frac{\hbar \omega_{GW}}{m_{Pl} c^{2}}
\ee
giving maximal density and maximal temperature, where $\omega_{GW}$ is the frequency of the gravitational waves, and the other symbols have their normal meaning (see section IX.F.1, List of definitions).  This spectrum holds down to those energies where the number density of resonant gravitons decreases even more steeply due to the separation of gravity from the other natural three forces, as one can show by solving the Boltzmann or Kompaneets equation on an FRW universe.

In four space-time dimensions the relativistic Boltzmann equation on the weak-gravity brane (for a detailed discussion see Appendix A) can be written in terms of a parameter $x$ which for convenience is defined as
\be 
x = \frac{h\,\nu_0}{k_B\,T_0}
\label{defx}
\ee
\noindent where $\nu_{0}$ is the frequency of a wave at emission, and $T_0$ is a temperature which is characteristic of the background source of energy.  In terms of $x$ the expression for ${\mathcal N}(\nu_0,t)$ is
\be 
\frac{\partial}{\partial t}{\mathcal N}(x,t) = \frac{ \kappa_{0}\,c}{k_B\,T_{0}}\frac{T_{g 0}}{T_0}\frac{1}{x}\frac{\partial}{\partial x}{\mathcal N}(x,t)\, ,
\ee
\noindent where $T_{g 0}$ is the initial graviton temperature, and $\kappa_{0}$ is the initial phase space integral of the matrix element squared for quadrupole emission of a graviton wave from the background.

 This expression can be put into a more compact form by defining the dimensionless parameter, $y$,
\be 
dy =  \frac{\kappa_{0}\,c}{k_B\,T_{0}}\frac{T_{g0}}{T_0}\,dt\, , 
\ee
or 
\be 
y = \int_0^t\,\frac{\kappa_{0}\,c}{k_B\,T_{0}}\frac{T_{g0}}{T_0}\, dt' 
\label{defy}
\ee
This gives a spectrum of, (compare with Eq. \ref{Planckseasp})
\begin{equation}
{\mathcal N}(x,y)\,=\,\frac{1}{\pi^2} \, \frac{x^{3}}{e^X - 1} 
\end{equation}
with 
\begin{equation}
X \, = \, \sqrt{\{x^2 + 2 y\}}
\end{equation}
where $y$ is a time-dependent term, not significant until the square of the dimensionless frequency $x$ matches $y$, so in the late universe the spectrum is steepened. As stated in the Introduction, this steepening coincides with the appearance of the Higgs boson.   For a detailed derivation of ${\mathcal N}(x,y)$ see Appendix A.

Invoking a resonance at energies below this implies a reduced density, and so will violate the surface density condition in the early universe, at the end of inflation.  Any attempt to resonate with energies beyond the Planck energy also fails.
The effect of $y$ on inflation is discussed in section VI. The time at which dark energy fails to get renewed is calculated in section IV.B.

\section{Creation equation}

\begin{figure}[h]
\centering
\includegraphics[bb=0cm 0cm 25.0cm 29.7cm,viewport=5.0cm 3.0cm 25.0cm
18.7cm,scale=0.5]{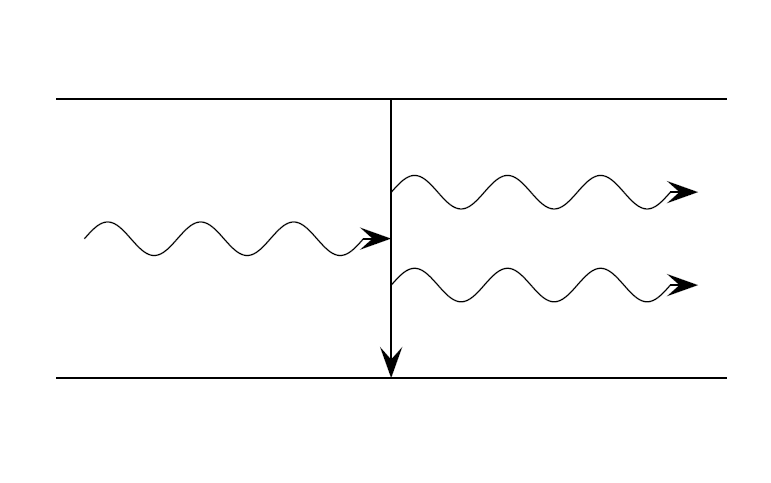}
\caption{The stimulated emission of photons}
\label{Stimulated_emission}
\end{figure}

In this section we describe the first new element of our model:  Dark energy implies a flow of energy from a background, which in our world may be considered as stimulated emission.  Stimulated emission scales as the square of the relevant phase space number density, and in our model we identify this phase space density in analogy to the holographic principle.   Figure \ref{Stimulated_emission} shows the process of coherent emission of gravitons from the background.  We picture the background as made up of particles at the highest energy conceivable, energies at which all natural forces combine; we will call these all-encompassing particles gravitons to emphasize the dominant nature of gravity.  Since gravitons are bosons, let us consider stimulated emission and determine the the expression for the number density for these bosons.  For photons, which are also bosons, the basic equation for stimulated emission is the Boltzmann equation describing the scattering of photons on electrons, in this application commonly referred to as the Kompaneets equation, e.g. \cite{ryb}.  For a discussion of the Kompaneets equation see Appendix B.

\begin{figure}[h]
\centering
\includegraphics[bb=0cm 0cm 25.0cm 29.7cm,viewport=5.0cm 3.0cm 35.0cm
18.7cm,scale=0.50]{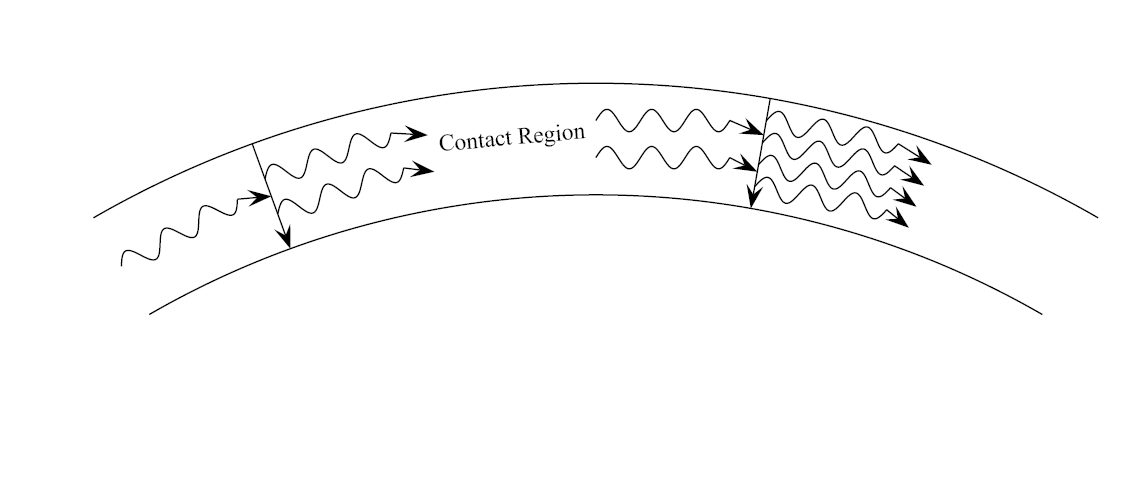}
\caption{The stimulated emission of gravitons in a shell}
\label{stimgrav}
\end{figure}

Whenever a black hole is formed, a large number of gravitons is generated, and we will argue below that it is plausible that most of these are produced in the very last collapse phase, just prior to making the black hole proper.  So a shell of gravitons, which we call the ``shell front'', is produced and propagates outwards as shown in Fig.\ref{stimgrav}.

In the following sections the Lorentz factors for the shell fronts and for black holes are derived, and an expression for the critical density is obtained.  These expressions are then used to determine the rate at which gravitational energy is created in our universe.

\subsection{Stimulated emission from the background gravity brane}

We propose a model in which energy is transferred from the strong-gravity (strong-gravity) brane (see Eq.(\ref{metric})) to our brane via resonance in gravitons, that is by stimulated emission.  Here we first give the microscopic derivation for all three cases, dark energy,  black holes, and inflation, before moving on to integrate the microscopic derivation into a cosmological context.

We are using the concept that in a 5D-space the phase space distribution is Planckian, with all the momenta in the fifth dimension being maximal, at Planck level.  This modifies the 4D phase space distribution such that the momentum in the fifth dimension acts like a chemical potential making the distribution essentially flat for all energies up to Planck level.  Interestingly, this implies that in such a case the Fermi and Bose statistics are very nearly the same in 4D-space.

The Planckian distribution in five dimensions is

\be
\frac{1}{e^{\{\hbar \omega/(k_B T_{Pl})\}} -1}
\ee
with
\be
(\hbar \omega)^2 \; = \; (c p_x)^2 + (c p_y)^2 + (c p_z)^2 + c^4 m_{Pl}^2\, ,
\ee
where the gravitons have a momentum with components $p_x$, $p_y$, and $p_z$.  So under all reasonable circumstances
\be
\frac{\hbar \omega}{k_B T_{Pl}} \, \simeq 1 \, .
\ee
The energy density of such a distribution is naturally of order $m_{Pl} c^2/l_{Pl}^3$.  We posit that this phase space distribution can never be exceeded.  This distribution is modified at low energies.   On the strong-gravity brane from the Higgs mass on down the distribution function is modified due to the rest mass of the Higgs.  This lowers the value of the distribution function.

We orient the coordinate system such that any motion or gradient of a potential is aligned with $z$.  We associate wavelengths via $\lambda_x = h/p_x$, $\lambda_y = h/p_y$, and $\lambda_z = h/p_z$.  We define the Lorentz factor $\Gamma_{Pl}$ with the motion which is required to bring the energy $c p_z$ up to Planck energy level.

We then assume that in the proper frame, to be defined below for each case, a maximal rate of energy transfer from the gravity brane to our brane is given by the rate gravitons in resonance move through the surface of the graviton, multiplied with the fraction of maximal momentum phase space.  

This maximal rate of energy transfer is modulated by the square of the coordinate space density on a surface in Planckian units. The transfer is derived from a rate at which a graviton on our brane becomes resonant with gravitons on the gravity brane, also on a surface.  The square of phase space density is obtained when producing a graviton each time, since we have an equal number of gravitons on the gravity brane and on our brane, all of which are in resonance.

At the end we reformulate this into some simple rules.

\subsection{The shell front}

In this section we describe the shell of gravitons emitted when a black hole forms and travels through space.  Further below we argue that the uncertainty in space and time implies a very large Lorentz factor for this shell front; the speed of the shell front is just below the speed of light.  This Lorentz factor compares the scale of the universe with the Planck scale and is given by
\be
\Gamma_D \; = \;  \frac{1}{2} \, \frac{r(z, z_{\star})}{l_{Pl}} \, \{H(z) \tau_{Pl}\}^{1/2} \, ,
\ee
where $H(z)$ is the Hubble parameter, and $r(z, z_{\star})$ is the spatial integral from redshift $z$ to $z_{\star}$
\be
r(z, z_{\star}) \; = \;  \int_{z}^{z_{\star}} \frac{c \, d z'}{ H(z')} \, .
\ee  
In concept this Lorentz factor is similar to the Gibbons-Hawking temperature.

This means that in the proper frame the $z$-component of the momentum is increased by a factor of $\Gamma_D$, and its length scale is reduced by the same factor.

We now consider the area of a graviton in the proper frame.  In the comoving frame of the shell front the rate of resonant gravitons going through can be written as a multiplication of several factors, the first of which is an emission transfer rate 
\be
\pi \lambda_{x} \, \lambda_{y} \, c \, \Gamma_{D} 
\ee
The speed of light appearing here is similar to a corresponding expression in radiative transfer, where emission can be written as energy density times $c/(4 \pi)$.  We will multiply with the energy density below.  The time step here is, however, the Planck time, and we can rewrite the speed of light as $l_{Pl}/\tau_{Pl}$ to obtain
\be
\pi \lambda_{x} \, \lambda_{y} \, l_{Pl} \, \Gamma_{D} \frac{1}{\tau_{Pl}}\, ,
\ee
which appears as a volume transfer rate due to the finite time step.
The second is the fraction of momentum phase space in resonance 
\be
\frac{l_{Pl}^2}{\pi \lambda_{x} \lambda_{y}} \, \frac{\Gamma_{D}}{\Gamma_{Pl}}\, .
\ee
The third is the energy density of the gravity brane 
\be
\frac{m_{Pl} c^2}{l_{Pl}^3}\, .
\ee
Transforming everything to observer frame takes out $\Gamma_{D}^2$, one for energy and one for rate.  The total transfer rate, which is obtained by multiplying all fo the terms above, gives an energy rate per patch of
\be
\frac{m_{Pl} c^2}{\Gamma_{Pl}} \, \frac{1}{\tau_{Pl}}\, .
\ee
Multiplying this with the number of patches and the Hubble time yields an estimate of the total energy transferred from the gravity brane to our brane per black hole.  This gives the maximum rate of energy transfer.

For this to reproduce dark energy as an energy transfer from the gravity brane this calculation shows that the extreme energy density on the gravity brane is required.  We will go through the numbers further below.

However, this effect is modulated by the proper phase space density squared to obtain a graviton, which we derive next.

\subsection{Critical phase space density}

The surface area of the front is $A$, and the number of gravitons is $N$. In terms of these quantities the surface density per Planck area, which is the filling fraction of gravitons, is 
\be
N \, \frac{ \pi l_{Pl}^2}{A} \, \Gamma_{D}\, .
\ee
This can be rewritten as,  following the line of reasoning above in analogy to radiative transfer,
\be
N \, \frac{ \pi l_{Pl}^3}{A } \, \frac{\Gamma_{D}}{\tau_{Pl}}\, .
\ee
We recognize that this measures the rate at which resonant gravitons go through the surface in the comoving frame per Planck volume.  This becomes the critical case when this rate reaches unity.  For stimulated emission we postulate that this is the critical phase space density and for the creation of a pair of gravitons this normalized phase space density must be squared.

This definition of criticality is closely related to the concept of the entropy of black holes, where also the surface density is of order unity \cite{bek2,bek3,bek4,bek5,bek6,bek8,mac1,mac2,pri,tho}.

We obtain a square here due to the fact, that we are using the same number of gravitons on our brane as on the gravity brane, which is analogous in the language of Feynman (vol. III, eq 4.28 on p. 4-8; \cite{FeynmanIII}) of having $n$ photons together with $n$ atoms. In the present case Feynman's atoms are analogous to the graviton distribution at momenta at or above the resonant graviton under consideration on the gravity brane, and the photons are analogous to the gravitons on our brane.  In our picture there is full resonance, so all gravitons on the shell front involved in the energy transfer are in resonance with gravitons on the gravity brane.  But much of momentum phase space on the gravity brane extends beyond the resonant gravitons, all at the limit of momentum phase space can bear.  The extremely small momentum phase space factor controls this (of order $10^{-105}$).

The number  of patches can be written similarly as
\be
\frac{A \; \Gamma_D}{\pi \, \lambda_x \, \lambda_y}\, .
\ee
This expression can be reformulated as the local rule that at most {\it one resonant pair of gravitons is created per patch ($\pi \lambda_x \lambda_y$) per Planck time in the proper frame}.  Transforming to the proper frame implies ($(\pi \lambda_x \lambda_y)/\Gamma_D$) is the size of the patch.  As noted above the single factor of $\Gamma_D$ arises from the condition that the argument derives from a rate.

By similar reasoning the surface density of gravitons is given by a surface density in the observer frame multiplied by the Lorentz factor $\Gamma_D$ to put it in the proper frame, since this similarly derives from a rate.

When we generalize specifically, we will add the required redshift factors.

\subsection{Black holes}

Black holes have a deep potential well, with a Lorentz factor of
\be
\Gamma_{BH} \; = \; \frac{M_{BH}}{m_{Pl}}\, .
\ee
The momentum $p_z$ goes up to Planck scale, but the lateral momenta remain unchanged.

The area for a single graviton is given by $\pi \lambda_x \lambda_y$, which in this case corresponds to the horizon of the black hole. The final result is that we obtain for the three-momentum space fraction, after combining the three coordinates,
\be
 \frac{l_{Pl}}{\lambda_{x}} \, \frac{l_{Pl}}{\lambda_{y}} \, \cdot 1 \, ,
\ee
which gives an emission rate in the observer frame whose order of magnitude is
\be
{\left(\frac{m_{Pl}}{M_{BH}}\right)}^2 \, \frac{m_{Pl} c^2}{\tau_{Pl}}\, .
\ee
One power of the black hole mass comes from the energy of the graviton in the observer frame and the other from converting the rate to the observer frame.

The surface density of gravitons is quasi-automatically critical for a black hole.  The entire surface is only one single patch.
Thus the same simple rule described above can be applied: at maximum {\it one resonant pair of gravitons is created per patch ($\pi \lambda_x \lambda_y$) per Planck time in the proper frame}.

\subsection{Inflation horizon shell}

For the inflation horizon shell the derivation of the Lorentz factor for the shell front differs from the derivation of the shell front Lorentz factor in Section 2 in that there is no time uncertainty  only a spatial uncertainty, because we can not say to better than a Planck length, where the horizon is. The details are given further below. In such a case, the Lorentz factor of the horizon is given by
\be
\Gamma_{infl} \; = \;  \frac{r_{infl}}{l_{Pl}}\, ,
\ee
which is a linear dependence.  At the end of inflation the value of $\Gamma_{infl}$ is about $10^{17.7}$.  This implies that in the horizon shell we pull up gravitons from a lower energy in the background by this factor.  At the end of inflation this is given by
\be
\Gamma_{infl} \, = \, \frac{\tau_{H} c }{l_{Pl} (1+z_{infl})^{3/2}}\, ,
\ee
On the left hand side we have
\be
\frac{m_{Pl} c^2 }{\Gamma_{infl}} \, , 
\ee
and on the right hand side we have
\be
\frac{c h}{\tau_H c} \, (1+z_{infl})^{3/2}\, .
\ee 
These two energies are identical, showing, that on the gravity-brane we reach down to the Higgs mass at exactly the same time that we reach the Higgs mass \cite{higgs,englert} at the observer location in the inflationary bubble.  The Higgs mass defines the end of inflation, since from the Higgs mass down on the gravity brane the distribution function is modified due to the rest mass of the Higgs.  This lowers the value of the distribution function, and so pulling up gravitons through the Lorentz factor of the horizon shell ceases.  The role of the Higgs particle in the inflation of the universe is discussed in more detail in section VIII.E.

Here again the surface density on the horizon is maximal anyway, and the patch scale is Planckian already.

Therefore the simple rule formulated above that at most {\it one resonant pair of gravitons is created per patch ($\pi \lambda_x \lambda_y$) per Planck time in the proper frame} holds here as well, so for all three cases.

\subsection{Rate of Energy Creation}

In the construction of the creation equation we assume that a local comoving frame exists in which the energy is transferred from the background.  We will show in the following section that a Lorentz factor of the comoving frame can be derived from the uncertainties in space and time.  Now write these rules as a function of redshift, so generalizing to the cosmological evolution.

The creation equation can then be written in the form of energy per unit time created in the form of resonant gravitons
\be
\frac{d E}{d \tau} \; = \; {\left( N_{GW}  \, \frac{\sigma_{Pl}}{4 \pi R_{s}^{2}} \, \Gamma_{D} \right)}^{2} \, \left( \frac{4 \pi R_{s}^{2}}{\pi \lambda^{2}} \Gamma_{D} \right) \, E_{GW} \Gamma_{E} \, \tau_{Pl}^{-1}
\label{creationeq}
\ee
in the comoving frame with proper time $\tau$ on a spherical surface of radius $R_s$ with graviton number $N_{GW}$ (An expression for $N_{GW}$ as a function of the mass, $M_{BH}$, of a black hole is given in section IV.B).  Here $\sigma_{Pl}$ is the Planck area, and $\tau_{Pl}$ is the Planck time (see section IX.F.1, List of definitions for the exact expression and the numerical value).  $\Gamma _D$ is a Lorentz factor applied to the surface density relative to the observer, and to the size of a patch in the comoving frame, and  $\Gamma_{E}$ the Lorentz factor as applied to the transformation of energy.  In the next section we show that for gravitons $\Gamma_{D} \, = \, \Gamma_{E} $.  $E_{GW}$ is the energy of the gravitons on the surface in the observer frame, and $\lambda \, \sim \, c h/E_{GW}$ is the corresponding length scale, with the numerical constant needed to make this relation an equality given by appropriate matching conditions.

This macroscopic expression is constructed to be similar to the nonlinear Boltzmann equation for stimulated emission in the relativistic case discussed in Appendix A, and also the Kompaneets equation describing stimulated emission in the non-relativistic case (see \cite{kom,zel,ryb}), but is really based on the microscopic derivation presented above.  The first term is in the form of an occupation number, and enters the relationship squared. It is the density of gravitons per Planck area in the frame of the surface.  The second term is the number of correlation patches in the surface in the observer frame, and $\Gamma_{S}$ is the Lorentz factor which transforms this number to the co-moving frame.  This Lorentz factor is necessary, since in the shell frame the correlation patch is smaller, and so a Lorentz factor $>> 1$ applies.  The third term is the energy in the observer frame, again transformed to the co-moving frame a Lorentz factor, and the last term is the rate in the surface frame.  To transform to the observer frame we take out the Lorentz boost of the graviton energy, and decrease the rate by the same factor.

The expression in Eq.\ref{creationeq} will be utilized in the following sections to describe the phenomena of dark energy, black holes, and inflation.

\section{Dark energy: Shell front}

\begin{figure}[h]
\centering
\includegraphics[bb=0cm 0cm 25.0cm 29.9cm,viewport=5.0cm 3cm 22.0cm
24.7cm,scale=0.32]{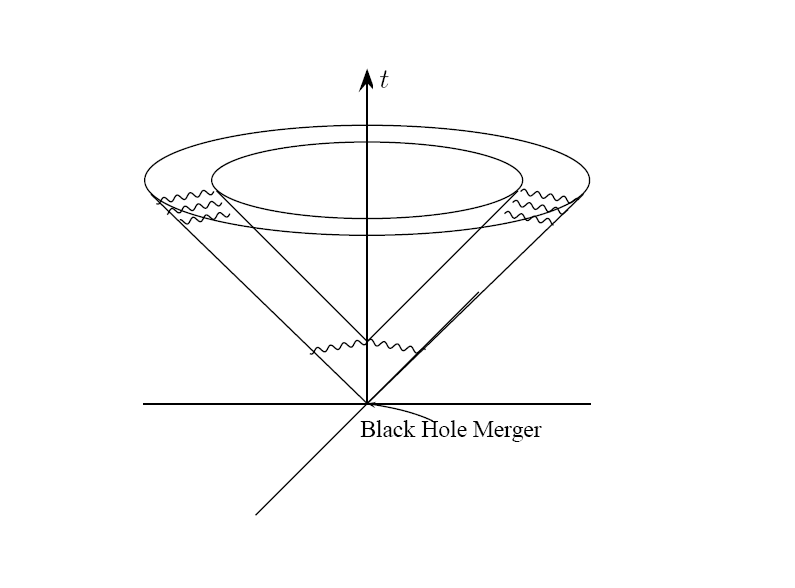}
\caption{Propagation of a shell front created by the formation of a black hole or the merger of two black holes.}
\label{merger}
\end{figure}

We have argued above and will discuss further in section V that in the last moments of black hole formation a shell front, consisting of many gravitons, is produced. To describe its motion, we derive  a Lorentz factor for the motion of the shell front given by  a combination of the spatial uncertainty $l_{Pl}/r(z, z_{\star})$ with the  Planck time uncertainty $\tau_{Pl} H(z)$ relative to the expansion rate of the universe; this is in a direct analogy to \cite{gibhaw77,gibhaw79} in which a combination of the scale of the universe with a Planck scale is used.  $r(z, z_{\star})$ is the distance scale of the universe, $z$ is the current redshift, and $z_{\star}$ is the redshift when the soliton front was born in the formation of a black hole, or analogously, in the merger of two black holes as in Fig.\ref{merger}.  
We emphasize that we are assuming that most of this energy emerges in the last few characteristic collapse times, so that the shell front has a thickness of just a few wavelengths in the observer frame, a characteristic which we will use.   There is a spatial uncertainty giving one Lorentz factor, and a temporal uncertainty giving another Lorentz factor. The combination of the two uncertainties decreases the effective Lorentz factor, since for an increase the front would be outside of resonance with the background.  These gravitons emitted into our world also have a quasi-Planck spectrum, with a characteristic temperature - given by the original black hole which creates the front - which decreases with the expansion of the universe.

\subsection{The Lorentz factor of the shell front}

In the following we use the analogy with the emission of a photon simultaneously with the emission of a particle of ultra high energy. After traversing a distance of the scale of the universe the time delay of the particle with respect to the photon is

\begin{equation}
\Delta t_{trav} \; = \; \frac{1}{2 \, \gamma_{UHECR}^2} \, t_{trav}
\end{equation}

\noindent where $\gamma_{UHECR}$ is the Lorentz factor of the particle, $t_{trav}$ the travel time, and $\Delta t_{trav}$ the time delay.  We will use the same expression for the delay below.

First we have to ascertain how precisely we can localize the shell front in the observer frame.  The shell front has a thickness of   several gravitational wavelengths, say $ 3 \, \varepsilon_{sh, 0.5}$
\begin{equation}
\lambda_{GW} \, = \, 3 \, \varepsilon_{sh, 0.5} \, l_{Pl} \, \frac{M_{BH}}{m_{Pl}} \, \frac{1 + z_{\star}}{1 + z} \, .
\end{equation}
However, since we have a very large number of gravitons, the precision to which we can measure length is in turn given by
\begin{equation}
\frac{\lambda_{GW}}{\sqrt{N_{GW}}} \, = \, \varepsilon_{sh, 0.5} \, \frac{3 \, l_{Pl}}{\sqrt{N_{GW}}} \, \frac{M_{BH}}{m_{Pl}} \, \frac{1 + z_{\star}}{1 + z} \, = \, \varepsilon_{sh, 0.5} \, \frac{3 \, l_{Pl}}{\sqrt{4 \pi}} \, \frac{1 + z}{1 + z_{\star}}
\label{lgw}
\end{equation}
which means that the precision is of order $l_{Pl}$.  It can obviously not be less than $l_{Pl}$, the limit is in fact exactly $l_{Pl}$.  For our adopted parameters $N_{GW} \, \simeq \, 10^{97}$.  In fact, this suggests that this equation misses a factor of $3/\sqrt{\{4 \pi\}}\,\simeq\,1$, which we will use below. This factor is similar to an equivalent factor of order unity in optical diffraction theory.

Any modification of the velocity of the shell front implies a speed below the speed of light.  The difference between this velocity and the speed of light can be derived using a random-walk analysis.  There are two uncertainties, one spatial and one temporal.
If the velocity differs from the speed of light, it can only be lower.  For any motion slower than the exact speed of light $c$ we have a speed $\beta c$, and the distance traversed towards the light front can be written as some distance $d$.  The spatial uncertainty of the distance could be positive or negative, but the distance is just slightly smaller than the distance to the light front.  $1-\beta$ can be written also as $1 -1/(2 \Gamma^{2})$ for $\Gamma >> 1$, where $\Gamma$ is related to $\beta$ as $\Gamma^{-2} \, = \; 1 - \beta^{2}$.  Since the velocity is $\beta c$, the advance is $\beta c \tau$, where $\tau$ corresponds to the age of the shell front, which is some time of order the age of the universe.  The relative delay in the distance travelled is 
\begin{equation}
\frac{1}{2 \Gamma^{2}} \, c \, \tau \, .
\end{equation}
The uncertainty in each step is the Planck length $l_{Pl}$. This spatial step-size can be either added or subtracted, and thus going from one time step to the next is equivalent to a one-step random walk.  Therefore the length squared $d^{2}$ changes on average as $d^{2} + l_{Pl}^{2}$, where $d$ is the total distance travelled.  Due to the one-step random walk, the uncertainty in the advance is 
\be
\frac{1}{2} \, \frac{l_{Pl}^{2}}{d^{2}} \, .
\ee
This implies by comparison
\begin{equation}
\frac{1}{2 {\Gamma_{d}}^{2}} \; = \; \frac{l_{Pl}^{2}}{2 d^{2}} \, .
\end{equation}

So we can identify the corresponding Lorentz factor as $\Gamma_{d} \, = \; d/l_{Pl}$.  We can identify $d$ with the distance scale, $r(z, z_{\star})$, in our universe  since we are considering the propagation of a light front (see section IX.F.2, List of definitions for a definition of $r(z, z_{\star})$). We are making a local comparison, so there is no extra factor of $(1+z_{\star})$. The Lorentz factor  in this case is  
\be
\Gamma_{d} \; = \;  \frac{r(z, z_{\star})}{l_{Pl}}\, .
\ee
In addition there is a corresponding uncertainty in time itself, which can also be written as a propagation slower than the exact speed of light.  However, time can go in only one direction, so we do not have to add squares. The temporal contribution to the Lorentz factor is determined from the probability that each step occurs.  This contribution is thus
\begin{equation}
\frac{1}{2 {\Gamma_{t}}^{2}} \; = \; \frac{H(z) \tau_{Pl}}{2} \, .
\end{equation}
Therefore we obtain 
\be
\Gamma_{t} \; = \; \{H(z) \tau_{Pl}\}^{-1/2}\, .
\ee
The final expression is obtained by combining these two Lorentz factors to obtain a common Lorentz factor for the shell front.  Adding the velocities would boost the Lorentz factor to the point where a normal graviton is outside resonance even with a Planck energy particle in the background Planck sea.  Therefore  the velocities should be subtracted, which by the usual rules of relativistic velocity transformation is equivalent to dividing the larger Lorentz factor by twice the smaller one.  The final Lorentz factor is
\be
\Gamma \; = \;  \frac{1}{2} \, \frac{r(z, z_{\star})}{l_{Pl}} \, \{H(z) \tau_{Pl}\}^{1/2} \, .
\ee
This is the Lorentz factor of the shell front in the frame of the freshly born black hole.  Typical numbers are $\Gamma \, \simeq \, 10^{31}$ for the Lorentz factor, which implies that the velocity of the shell front is $1-10^{-62.3}$ times the speed of light.

Extrapolating the behavior of the Lorentz factor into the future yields an asymptotic constant, as $H(z)$ approaches a constant for $(1+z) \, -> \, 0$, and so $r(z, z_{\star})$ (see section IX.F.2, List of definitions) also approaches a constant.

This derivation should be applicable to any signal, that is, to any change of a wave or massless particle travelling with the speed of light.  A change in the wave or an increase or decrease in the flux of particles constitutes a signal and thus corresponds to new information.  The derivation shows that any such signal always travels at a velocity just below the speed of light $< c$, just like a particle with mass.  However, then we have to ascertain again, whether we can actually reach the Planck limit precision, and the answer will generally be no.  Therefore the limit comes down to the wavelength of the energy packet divided by the square-root of the number of coherent such packets $N$; this clearly says that this argument works only for Bosons, not for Fermions.  This gives for the uncertainty, using the matching condition above which requires the extra factor of $\sqrt{4 \pi}$
\be
l_{min} \, = \, \frac{c h}{E_{1}} \, \sqrt{\frac{4 \pi}{N} }\, ,
\ee
where $N$ is the number of Bosons of individual energy $E_{1}$ which act coherently.  The total energy is then $E_{\Sigma} = N E_{1}$, and we obtain for the Lorentz factor
\be
\Gamma_{r} \; = \; {\left( \frac{d}{8 c h} \, {\left( \frac{E_{1} \, E_{\Sigma}}{\pi}  \right)}^{1/2} \right)}^{1/2} \, .
\label{coherenceeq}
\ee

This should also hold for sub-atomic scales, but it requires a very large number of coherent Bosons to be of any interest.  We will explore the consequences which may arise from this result elsewhere, but will point out one consequence of interest in the last section of the paper, section VIII.  For a single wave, this energy corresponds to a mass of $\sim \; 10^{-84.3}$ g, or, for all of the waves in one soliton, to a mass $\sim \; 10^{97}$ times higher, $\sim \; 10^{12.6}$ g, numbers extremely small compared to the mass of the black hole just born, which is $10^{6.5} \, M_{\odot} \, = \, 10^{39.8} \, {\rm g}$ using the numbers adopted.

\subsection{The Shell front}

Using the Lorentz factor derived for the shell front, we can work out the rate of dark energy production.  We have verified  that the occupation number in Eq.(\ref{creationeq}) (for an exact definition of this number see Eq.(51)) is close to unity for our adopted parameters. For more massive black holes it would be even larger; for significantly lower mass black holes it would be much less than unity, and so these black holes would not contribute significantly to dark energy production.

Inserting the number of gravitons $N_{GW}$ given by the creation of a black hole or the merger of two black holes, into Eq.\ref{lgw} above we obtain (spin $J = 0$ approximation)
\begin{equation}
N_{GW} \; = \; 4 \pi \; {\left( \frac{M_{BH}}{m_{Pl}} \right)}^2 \, {\left(Ê\frac{1 + z_{\star}}{1 + z} \right)}^{4}\, .
\label{ngw}
\end{equation}
Ideally we would integrate the creation equation, and self-consistently derive the time dependent dark energy contribution.  Since this is beyond the scope of the present investigation, we simply determine the power-law dependencies.  The exponent of the redshift term follows from the requirement that we obtain self-consistency at the end.  No other simple power-law dependence would allow this.  This approach should be considered as a consistency check, since we do not propose a more general equation, and then derive the specific expression.  However, we show below that dark energy does end, far in the future of the universe, so postulating this expression goes beyond reproducing known observations.  The first term in the creation equation (Eq. \ref{creationeq}) is the ``occupation number'' $N_{occ}$ squared,
\be
N_{occ} \; = \; {\left( \frac{N_{GW} \sigma_{Pl}}{4 \pi d_{L}^{2}} \, \Gamma_{D} \right)}\, .
\ee
This number is of order unity (using our adopted parameters it is $N_{occ} \, \simeq \, 10^{0.5}$) and for $\Gamma_D\, =\, \Gamma$ it is ``critical" as required by our approach. For the definition of the luminosity distance $d_L$ see section IX.F.2, List of definitions.

For the Schwarzschild metric the scale of the resonant correlation patch is given by
\be
\lambda \; = \; \frac{1}{2} \, {\left(\frac{\pi}{3}\right)}^{1/2} \; \frac{G_N M_{BH}}{c^{2}} \, \frac{1 + z_{\star}}{1 + z} 
\label{corrp}
\ee
where we will justify the numerical factor $(\pi/3)^{1/2}$ from a matching condition below.  The factor of 1/2 is appropriate in the case that we identify the diameter of the Planck shell with a half-wavelength, that is using adjacent nodes of a wave.  A numerical value is $\lambda \, \simeq \, 10^{13}$ cm, again using our adopted parameters.  The number of correlation patches on the shell front in the comoving frame is about $10^{66.4}$, using our adopted parameter values.

The resonant energy of a graviton is
\be
E_{GW} \; = \; \frac{1}{8 \pi} \; m_{Pl} c^{2} \frac{m_{Pl}}{M_{BH}} \, \frac{1 + z}{1 + z_{\star}}\, .
\label{egw}
\ee
In the comoving frame  (i.e. after multiplying with $\Gamma$) this is about $10^{+1}$ erg.
Finally, using the expression for the Lorentz factor from above and taking out the factor $\Gamma^{2}$ to move to the observer frame, Eq.\ref{creationeq} becomes,
\be
\frac{d E_{T}}{d t} \; = \; \frac{3}{2} \, M_{BH} \, c^{2} \, {\left(Ê\frac{1 + z_{\star}}{1 + z} \right)}^{3} H(z)\, .
\label{eq1}
\ee
Inserting all the expressions then gives the redshift dependence of dark energy, once folded with the density of black holes, and their redshift dependence.

We can work out directly the required rate of energy production, and then match the result to Eq.\ref{eq1} to obtain the values of the numerical factors in our approximations.  What we need per black hole is given by 
\begin{equation}
\frac{d}{d t} {\left( N_{GW} E_{GW} \right)}\, ,
\label{doteqngw}
\end{equation}
which is 
\begin{equation}
\frac{3}{2}  \, M_{BH} \, c^{2} \, {\left(Ê\frac{1 + z_{\star}}{1 + z} \right)}^{3} H(z)\, . 
\label{eq2}
\end{equation}
Matching the expressions in Eqs. \ref{doteqngw} and \ref{eq2},  and using the expression for $\lambda_{GW}$ in Eq.\ref{corrp} the factor is found to be $(\pi/3)^{1/2} \, \simeq 1.02$.
Setting $ N_{GW} E_{GW} \, \sim \, \rho_{DE}$, we see that the energy source in our universe scales with $P_{DE} H(z)$. We will use this below when we will employ the energy-momentum tensor and the Einstein equations.

It is straight-forward to verify that a shell front cannot just collapse into mini-black holes. Such a front does not have the critical density to do this during its formation and never reaches the critical density.  As it expands its density in its own frame scales with $\Gamma^{-1}$, since the scale of the radius enters the Lorentz factor.  At the same time its thickness scales also with $\Gamma^{-1}$.  So the two time scales never cross.

Combining this with the observed space density of supermassive black holes today cancels all redshift dependence and gives a numerical value for dark energy, which is consistent with observation to within the rather large error bars, as shown in the next subsection.

For a more detailed discussion of the conditions at the end of inflation see Appendix C and D.

\subsection{Observations}

The relations derived in the previous section for the rate of dark energy production can now be used to obtain a value for the dark energy density, which can be prepared to the observed result.  For simplicity we use the same values of the parameters as in Section IV.A.  However, we initially keep these parameters arbitrary in the expressions in order to show the parameter dependencies.

When a black hole is created or two black holes merge, the maximum efficiency of converting rest-mass into energy is 50 percent.  This efficiency is reached when either two black holes of opposite, maximal spin and of equal mass merge \cite{haw1}, or when a black hole forms out of a collapse into a non-rotating black hole \cite{wal}.  We use this limit here and address the cases of rotating black holes, orbital spin and mergers of black holes of unequal mass in a future investigation.

The total energy is the number of gravitons per black hole times the energy of each graviton times the density of black holes 
\be
\rho_{DE} \; = \; 4 \pi {\left( \frac{M_{BH}}{m_{Pl}} \right)}^{2} \, {\left( \frac{1 + z_{\star}}{1 + z} \right)}^{4}  \cdot \frac{1}{8 \pi} \, {\left( \frac{1+z}{1+z_{\star}} \cdot  \frac{ m_{Pl}^{2} c^{2}}{ M_{BH}} \right)} \, n_{BH, 0} \, (1+z)^{3}\, .
\ee

To obtain an estimate of the dark energy density we set  $M_{BH} = 3 \cdot 10^{6} \, M_{\odot}$ and $z_{\star} = 50$; The density of black holes at creation, using today's value, is $n_{BH, 0} = 10^{-2.2} Mpc^{-3}$ \cite{car}.  The original density of super-massive black holes could be substantially higher, e.g. by a factor of order 8, if black holes grow mostly by merging (e.g. \cite{berti,ger}).  In addition, the observational uncertainty in the number itself is known to be a factor of 2.5 (one sigma), see \cite{car}.  The black hole mass could perhaps be $10^{7} \, M_{\odot}$, giving another factor of 3.   Furthermore, the creation redshift $z_{\star}$ could be larger or lower than our adopted example of 50.  Using the estimate in \cite{bie} of the first stars forming at redshift 80 at most, super-massive stars seem unlikely to form earlier than at a redshift of 70, since the difference corresponds to a few million years, which is the life-time of very massive stars.  This would modify the number by a factor of 4.  In view of all these uncertainties  the parameters allow the match with dark energy of $10^{-8.0} \, erg/cm^{3}$; using today's parameters naively without any correction yields a value for the density two orders of magnitude lower; this also matches what is usually called the entropy of the universe.  Since the largest correction here derives from assuming that super-massive black holes grow by merging with each other, this seems to be an implied consequence.  We do not use the black hole density in the creation equation; we follow the evolution of the shell front of a single black hole. In order to obtain the observed vacuum energy density today the original number of black holes would have to be $n_{BH, 0} = 10^{-0.1} \, M_{BH, 6.5}^{-1} \, {\rm Mpc^{-3}}$, scaled to today.  This value assumes that all other parameters are unchanged. In the total energy density the dark energy contribution scales with the total mass in super-massive black holes.  Dark energy has grown by a factor of about $(1+z_{\star})^{3} \, \simeq \, 10^{5}$ since it was first created.

One check here is that the efficiency of turning normal matter into the first super-massive black holes should be small \cite{silk}.  We use the numbers of $n_{BH, 0} = 10^{-0.1} \, M_{BH, 6.5}^{-1} \, {\rm Mpc^{-3}}$ as the original black hole density scaled to today, and consider a black hole of $3 \cdot 10^{6} \, M_{\odot}$ made out of baryonic matter.  This estimate uses for comparison the entire baryonic mass out to the next sphere of influence of another similar black hole.  At the adopted formation redshift of $z_{\star} \, = \, 50$ this sphere has a radial scale of about $10^{1/3} (1+ z_{\star})^{-1} \, {\rm Mpc} \, = \, 21 \, {\rm kpc}$.  Since the specific volume scales linearly with the mass of the black hole,  dividing the black hole mass by the specific volume renders the efficiency independent of the mass.  Furthermore, the efficiency is also independent of the formation redshift adopted, since we use numbers referring to today.  This gives an efficiency of $3 \, 10^{-5 \pm 0.4}$, which, considering the original observational uncertainties \cite{car}, matches the extremely low efficiency found elsewhere.

\section{Black holes: Planck Shell}

When a black hole is created (e.g. \cite{bar2,bek1,haw4,wal}), or when two black holes merge  \cite{haw1}, we argue that all of the rest mass energy can go into the black hole and partially emerge as gravitational waves, neutrinos or electromagnetic emission.  The limiting case is that half of the energy goes into gravitons. This limit is  valid for a Schwarzschild metric, and also for two equal mass black holes of maximal and opposite spin in the plane of rotation merging (see also \cite{coo}).  Although our metric is not Schwarzschild, we nevertheless assume the same division of energy into gravitational waves and neutrinos or electromagnetic emission.  

This can be derived from our creation equation Eq.\ref{creationeq}.  Consider the collapse of a massive configuration towards a black hole; this implies that we need to include the Lorentz factor corresponding to the potential depth, and further below we will argue that this is $M_{BH}/m_{Pl}$. This necessarily corresponds to a non-vacuum solution.   We also want to get the total amount of energy coming out in the burst, and so we integrate in the observer frame by multiplying with $\tau_{Pl}$. We note that at the moment when the black hole is born, the first term in the creation equation is unity times $\Gamma_D^{2}$, the second term is unity times $4 \, \Gamma_{D}$, the the third term is the energy of a gravitational wave in the observer frame, $E_{GW} \sim \, M_{BH}^{-1}$ times $\Gamma_{E}$.  All this is in the moving frame of the collapsing black holes. Here as seen from an observer $\Gamma_D \, = 1$, but $\Gamma_{E} \, = \; M_{BH}/m_{Pl}$.  The total energy coming out (over $\tau_{Pl} \times \Gamma_E$) is
\begin{equation}
E_{BH} \; \simeq \;  \, M_{BH} \, c^{2}\, ,
\end{equation}
however, for convenience we assume in the following that the numerical factor in our simple approximation is just 1/2, and not unity.  This supports the notion, that at the formation of a black hole a large number of gravitons are emitted, matching in total of order half the rest mass energy of the black hole mass, seen from a distant observer.

However, this entire argument is more of a consistency check again, since the actual value of the emitted energy is obtained by integrating over the progression of the Planck shell, as it first forms, then grows in the collapse of a massive object, see \cite{joshi12}. At each step when the Planck shell has enlarged by a Planck area in its own frame, it can emit a graviton per Planck time.  This is rather similar to the initial formation and subsequent growth of a shock wave in non-stationary gas dynamics, which also separates causal regimes, \cite{CourFried48}.

We will deal with black hole spin elsewhere.  As a guiding principle in the following we adopt this limiting efficiency for the formation of the first black holes, and assume all of the radiated energy to be available for gravitons. At very high particle energies we take the graviton to incorporate all natural forces.  The implication is that at the thermodynamic limit gravitons are generated.  We identify what is normally called ``entropy'' of black holes with gravitons, both far from the black hole, and also just outside what is normally referred to as the horizon.  It is this ``shell front'' of gravitons propagating through the universe from each black hole formed (or from each black hole merger) which  provides the connection of black holes to the cosmos.  However, the energy of the gravitons depends on the black hole mass, so the energy of these gravitons will generally be different for different black holes.  This is also different from current approaches (e.g. \cite{bou}), where these two populations of gravitons do not exist.

\subsection{Planck shell}

\begin{figure}[h]
\centering
\includegraphics[bb=0cm 0cm 25.0cm 29.7cm,viewport=5.0cm 1.0cm 20.0cm
18.7cm,scale=0.50]{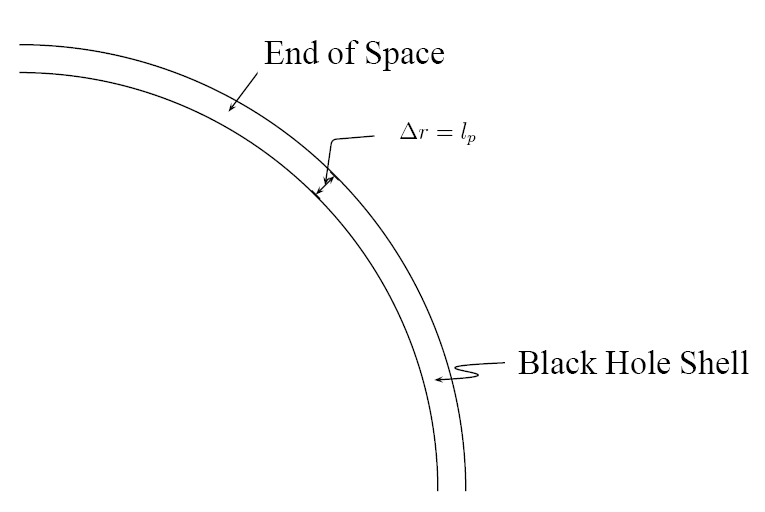}
\caption{The Planck shell for a black hole}
\label{blckshll}
\end{figure}

We propose that a black hole can be pictured as a ``Planck shell" full of gravitons at potential depth $M_{BH}/m_{Pl}$ in terms of a Lorentz factor.  This is rather similar to the concept of a ``stretched horizon" \cite{pri,sus}.  This shell is an impenetrable barrier, and the ensemble of these gravitons constitutes the black hole as shown in Fig.\ref{blckshll}.  Space-time ends at the inner surface of the shell; there is no physical meaning to space-time coordinates in the region enclosed by the shell.    This concept was also later suggested in terms of a ``firewall" by Almheiri et al. \cite{firewall1}.  

These gravitons have a quasi-Planck spectrum.  The energy density on the Planck shell, $\varepsilon_{BH}$, is given by the Planck limit multiplied by the ratio of the Planck mass to the black hole mass
\be
\varepsilon_{BH} \; = \; \frac{m_{Pl}}{M_{BH}} \, \frac{m_{Pl} c^2}{4 \pi l_{Pl}^3}
\ee
 The creation equation above requires that energy be pulled from the background, so the black hole does not evaporate but very slowly increases in mass.

The derivation of the Lorentz factor in terms of the black hole mass begins with the gravitational potential.  The invariant line element for a curved space-time background can be expressed in terms of the gravitational potential, $\Phi(\vec{r})$, as
\be 
ds^2 \; = \; -{\rm e}^{2\,\Phi(\vec{r})/c^2}\,c^2\,\,dt^2 + g_{ij}\,dx^i\,dx^j \nonumber
\ee
The ratio of the frequency of radiation emitted at a point, $\vec{r_S}$, at the Planck shell surface to that at a distant observation point, $\vec{r_O}$, is given by
\be 
\frac{\nu_O}{\nu_S} \; = \; \frac{\sqrt{(g_{00}(\vec{r}_{S}))}}{\sqrt{(g_{00}(\vec{r}_{O}))}}
\label{frequencyratio}
\ee
The Lorentz factor $\Gamma_{BH}$ is related to the frequency ratio in eq. \ref{frequencyratio} by the expression
\be
\frac{\nu_S}{\nu_O} \; = \; \Gamma_{BH}
\ee
and $\Gamma_{BH}$ is related to the $g_{00}(r)$ component of the metric tensor by the relation given above. The form of $\Gamma_{BH}$ which gives consistency can be written as
\be
\Gamma_{BH} \; = \; f(r) \, (r - r_S)
\ee
where $f(r)$ is defined by the requirement that $\Gamma_{BH}$ has the proper values at infinity and at the horizon. We use the ansatz
\be
f(r) \; = \; \frac{f_0}{r} \; .
\ee
This results in
\be
f_0 \; = \; \frac{r_S + l_{Pl}}{l_{Pl}}
\ee
The $g_{00}$ element of our metric tensor element then has the form
\be
g_{00} \; = \; {\left( 1 - \frac{r_S}{ 2 r}\right)}^{2} \, ,
\ee
and the invariant line element is 
\be 
ds^2= -(1-\frac{r_S}{2\,r})^2\,dt^2 + \frac{1}{(1-\frac{r_S}{2\,r})^2}\,dr^2 +r^2\,d\theta^2 + r^2\,\sin^2(\theta)\,d\phi^2 \, .
\label{toy}
\ee
This results in a Lorentz factor of the required form
\be 
\Gamma_{BH} = \frac{r_S}{2\, l_{pl}}\, = \, \frac{M_{BH}}{ m_{pl}}\, . \nonumber
\ee
This metric is of the extreme Reissner-Nordstr\"{o}m type with the charge $Q$ obeying
\be
G_N Q^2 \, = \, {\left(G_N\,M/c^2\right)}^2 \, . 
\ee

We are not proposing the metric in Eq. \ref{toy} as the correct form for the metric at a Planck surface.  We want to show only that a Lorentz factor of the required form can be obtained for non-vacuum solutions of the Einstein equations.  The actual form of the metric near the Planck surface is of course determined by the interaction of the surface with the background energy sea and with any other fields which may be present.  The determination of the true metric is beyond the scope of the present investigation but will be the subject of a future investigation. 

The filling condition of the occupation number for the Planck shell is obviously satisfied ($\Gamma_D \, = \, 1$ and $\Gamma_E \, = \, M_{BH}/m_{Pl}$).  In resonance the graviton on the Planck shell is exactly the Planck energy in the Planck sea, the background.  We also note, that laterally the graviton has the size of the circumference of the black hole horizon, and radially the thickness is the Planck length.  A similar model with an extreme anisotropy was investigated by t'Hooft, \cite{thooft}.  If all of the gravitons on the Planck shell are in a coherent state, i.e. all of the gravitons are aligned, our model reproduces the spin of a Kerr black hole.

Using the creation equation, the occupation number can be shown to be of order unity independent of our standard parameter choices. Creating one Planck graviton per Planck time on the Planck shell implies that in the observer frame the luminosity is locally
\be
L \; = \; \frac{1}{2} \, \frac{m_{Pl} c^{2}}{\tau_{Pl}} \, {\left(\frac{m_{Pl}}{M_{BH}}\right)}^{2} \; .
\ee
The factor of $1/2$ is an approximation as explained in section IV.D.
Half the energy obtained from the background field is radiated away and half is absorbed, increasing the mass of the black hole.  There is no net evaporation of black holes in this approach, although the Hawking process still happens \cite{haw3,pag1,pag2,pag3}.  However, the Hawking process does not drive the evolution of the black hole, it is actually much weaker than the process described here, the interaction with the background.  This implies that primordial black holes all grow to the point where their growth time scale becomes the age of the universe.  This allows rather strong limits on the initial number of primordial black holes \cite{haw0,carr75,san78}, as we will discuss further below.

The typical energy of the graviton coming out in the observer frame is given by
\be
E_{o.f.} \; = \; \frac{1}{8 \pi} \, m_{Pl} c^{2} \, {\left(\frac{m_{Pl}}{M_{BH}}\right)} \, .
\ee
To a distant observer this is measured as a redshifted Planck spectrum.

A black hole forms, when a Planck shell is created at some finite radius in a collapsing configuration.  Therefore we have to question whether at that point we introduce a new variable in terms of how many gravitons get generated as a function of the enclosed mass at the moment at which  the Planck shell is created.  However, since in the end the black hole cannot have any extra parameters, such as this specific radius, the graviton generation must be independent of any extra parameters. A fortiori graviton generation must be the same as if the black hole were to start growing from a Planck mass.  The first formation of a Planck shell, and its subsequent growth are just a function of its mass.

\subsection{Graviton emission rate in an expanding universe}

The rate at which gravitons are emitted from a black hole in an expanding universe without dark energy can be calculated in a semiclassical approximation.
Following Parikh and Wilczek \cite{parikh} we calculate the emission rate for a graviton of frequency $\omega$ by calculating the action in the WKB approximation for a black hole in an expanding universe.
The graviton emission probability for a black hole in an expanding universe can be calculated from the McVittie metric \cite{mcvittie}
\be 
ds^2 = \frac{-(1-\frac{\mu(t)}{2\,r})^2}{(1+\frac{\mu(t)}{2\,r})^2}\,dt^2\,+\,e^{\beta(t)}\,\left(1+\frac{\mu(t)}{2\,r}\right)^4\left[ dr^2\,+\,r^2\,d\Omega_2\right]
\label{met}
\ee
where
\be 
\frac{\dot{\mu}(t)}{\mu(t)} = -\frac{1}{2}\dot{\beta}(t) \, .
\label{mu}
\ee
The semiclassical emission rate is thus proportional to (for a more detailed derivation of this see Appendix E)
\be 
\Gamma_{em} \sim e^{-2\,\Im S} \,=\, e^{-4\,\pi\,a(t)^2\,\omega\,(5\,m - {\omega}/{2})}
\label{Gamma}
\ee
Neglecting the term quadratic in $\omega$, the exponential in Eq.(\ref{Gamma}) becomes a Boltzmann-like factor with a temperature which decreases with increasing time.  This is consistent with our model of ever-cooling black holes.  The rate expression in Eq.(\ref{Gamma}) does not take into account the cooling due to the interaction of the Planck surface with the background.  This interaction creates further growth, i.e. cooling, of a black hole.

\subsection{Accreting black holes}

When black holes accrete, they do not pull any additional energy from the background; however, the accreting matter produces a mix of electromagnetic emission, neutrino emission, and graviton emission in addition to allowing the black hole to grow.  The graviton component of the emission is usually dominant, and produces a general gravitational wave background, which may be just below the current upper limits  \cite{lig}.  This additional gravitational wave emission, which is some factor of order $< \, 10$ above the electromagnetic emission, provides a universal background of gravitational waves.  

We can vary our creation equation by assuming that a particle, say a proton, accretes and gains energy by the potential depth, thus gaining a factor of $M_{BH}/m_{Pl}$ (using zero redshift for simplicity).  The energy will rapidly exceed the Planck energy, creating additional Planck energy gravitons.  We reiterate that for a black hole, both the occupation number and the number of correlation patches are unity.  This implies that the output becomes
($\epsilon$ is the fraction of the accreted energy which is re-radiated)
\begin{equation}
\frac{d E}{d t} \; = \; \left( \epsilon \, \frac{m_{Pl} \, c^{2}}{\tau_{Pl}} \, \frac{m_{Pl}}{M_{BH}} \right) \, \times \, \left( \frac{\dot{M}_{BH} \, \tau_{Pl}}{m_{Pl}} \, \frac{M_{BH}}{m_{Pl}} \right)
\end{equation}
where the first term arises from the normal production of gravitons, and the second term arises from the infall of matter, leading to the splitting of the energy into Planck level gravitons, and the resulting multiplicity.  We choose $\epsilon$ to be one-half to describe the outgoing graviton emission in our simple approximation.  This equation then reduces to the trivial relationship
\begin{equation}
\frac{d E}{d t} \; = \; \epsilon \, \dot{M}_{BH} c^{2}
\end{equation}
Baryon number, charge and lepton number are conserved in the creation of new particles, but baryons and leptons constitute only a minute fraction of all the emitted and absorbed particles.

In \cite{car}  the energy density of such a background was estimated. The mass of black holes contributing to the energy density of the universe can be written as
\begin{equation}
\Omega_{BH} \, = \;   2 \cdot 10^{-6 \pm 0.40}
\label{obh} 
\end{equation}
Assuming that energy is emitted only during accretion, and that growth of super-massive black holes is mostly via this accretion \cite{sha,nov}, the energy input into intergalactic space is given by $6 \cdot 10^{-15 \pm 0.40} \, \epsilon_{0.3} \, erg \,  cm^{-3}$, using an accretion efficiency of 0.3 to produce gravitational waves. For a Schwarzschild metric the maximal number is 0.5, with about 0.06 going into electromagnetic emission. So the maximum fraction of the energy which can appear as gravitational waves is 0.44.  For a Kerr metric the electromagnetic part could reach 0.43, so the fraction of energy appearing as gravitational waves could be very much less \cite{bar1}.  

In terms of critical energy density the density in Eq.\ref{obh} transforms to
\begin{equation}
\Omega_{GW,SBH} \, = \; 6 \cdot 10^{-7 \pm 0.40} \, \epsilon_{0.3} \, f_{eff} \, {\left( {1+z_{acc}} \right)}^{-4}
\end{equation}
with the index $GW,SBH$ signifying the gravitational wave emission due to accretion by super-massive black holes. $f_{eff}$ is the fraction of the final mass that is due to accretion after formation, and  $z_{acc}$ is the redshift for the epoch, when most black hole accretion occurred.  If black hole growth is mostly due to accretion, the $f_{eff}$ will be very close to unity; if black hole growth is mostly by black hole mergers, then obviously $f_{eff} << 1$ is possible.  From cosmic evolution studies, the greatest merger activity is known to be somewhere between redshift 1.5 and 2 approximately, so the redshift term may contribute another order of magnitude.  There will also be a corresponding contribution due to accretion by stellar mass black holes.
\par
In the following sections we note again that black holes probably grow by merging with each other, and in that case the gravitational wave background estimated here may be an order of magnitude lower.  This gravitational wave background is  just the original gravitational wave production; it should not be confused with the dark energy background, whose energy density is of order $10^5$ higher than that of the background estimated in this section.  
\par
Since we are not proposing a specific model for the production of this energy density as a function of black hole mass and cosmic epoch, we refrain from proposing a spectrum for the associated gravitational waves. However a reasonable assumption is that the energy density spectrum reflects the mass function of super-massive black holes. Thus it may be quite low at the frequencies associated with $10^{9} \, M_{\odot}$ black holes, and peak at the frequencies associated with $10^{6.5} \, M_{\odot}$ black holes, shifted by the redshift dependence of the accretion history.  This specific number of $10^{6.5} \, M_{\odot}$ is uncertain by possibly a factor of 3, as already noted.

\section{Inflation: Horizon shell}

The model for graviton creation described in the preceding sections may also explain the inflationary period \cite{guth81,linde81,linde82}.  During inflation new gravitons are produced at the horizon in a ``horizon shell", provided that the newly created gravitons lose their energy adiabatically with a time scale proportional to the Planck time, independent of redshift within the inflationary period.  Exponential behavior is common in any approach to inflation. However, in our model the structure of the universe during its inflationary period is the one seen by an observer. The inflationary period stops when the new gravitons pulled from the background reach the particle energy where the background fails to provide enough gravitons.  In this concept the universe starts with one Planck-energy graviton in one Planck volume.  This kind of speculation has been explored for many years, e.g. by Lemaitre \cite{lem}.

The concept is that near the horizon we also have a shell, which is at maximum energy and particle density. Thus it satisfies the filling condition.  

Within the scenario above the creation of new gravitons always occurs inside our universe.  Each new graviton which is produced stretches the wavelength of the gravitons produced prior to the new graviton.  This behavior can be diagrammed as

\begin{figure}[h]
\centering
\includegraphics[bb=0cm 0cm 25.0cm 25.7cm,viewport=4cm 15cm 18.0cm
26.7cm,clip,scale=0.8]{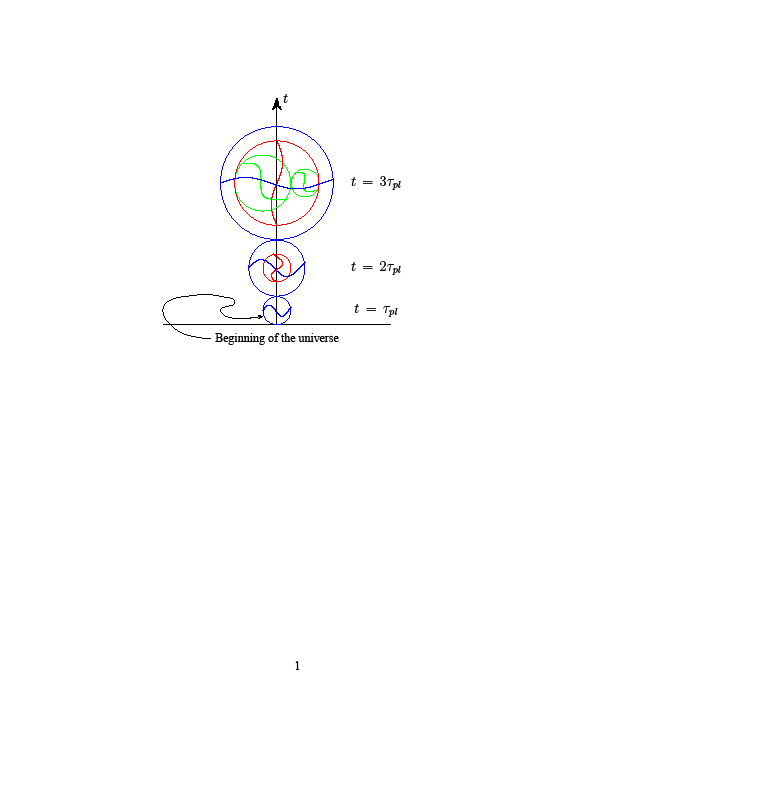}
\caption{The early growth of the universe starting from a single Planck cell.  Each cell stimulates an additional cell of the same energy.  The wavelengths of the older gravitons (cells) are stretched by the addition of new gravitons.}
\label{Inflation}
\end{figure}

If the universe has a size given by 

\begin{equation}
\Delta x_{n}' \, = \, N_{n} \, l_{Pl}
\end{equation}

\noindent where 

\begin{equation}
N_{n} \, = \, N_{0} \, R(t_{n}) \, .
\end{equation}

\noindent where $N_{0} \; = \; 1$ in our model, and this equation represents a 1-D slice through the early universe.

From the diagram the expression for $R(t)$ is seen to be

\begin{equation}
R(t_{n}) \; = \; N_{0} \, 2^{n} \;\;\; n \; = \; \frac{t_{n}}{\tau_{Pl}}
\end{equation}

and

\begin{equation}
R(t_{n}) \; = N_{0} \, e^{n \ln 2} \; = \; N_{0} \, e^{(t_{n}/\tau_{Pl}) \ln 2}
\end{equation}

Thus the initial wavelength

\begin{equation}
\lambda_{0} \; = \; l_{pl}
\end{equation}

will increase as more and more gravitons are produced, so that

\begin{equation}
\lambda_{n} \; = \; R(t_{n}) \, \lambda_{0}
\end{equation}

This of course is an approximation because at some point the wavelength of each graviton becomes too large and the frequency becomes to small for the graviton to stimulate emission of new gravitons at the peak energy of $m_{Pl} c^{2}$.  The older gravitons generate new gravitons at the lower frequency and therefore with lower luminosity.

This is then directly the normal exponential growth

\be
r \; = \; r_0 e^{t/\tau_{infl}}\, ,
\label{req}
\ee
where we identify $r_0$ with the Planck length $l_{Pl}$, and $\tau_{infl}$ with the Planck time $\tau_{Pl}$ to within a factor of order unity. As shown earlier for the shell front, this then gives a Lorentz factor  of 
\be
\Gamma_{infl} \; = \; \{ r/l_{Pl} \} \; = \; e^{t/\{\tau_{infl}\}}\, .
\label{inflation}
\ee
Therefore the apparent velocity of the horizon increases exponentially, a velocity which we translate into a Lorentz factor.   A shell front with this Lorentz factor, interacting with the background will resonate in the background with particles with lower energy, and therefore at lower phase space density.  The contributions to the counting of the powers of the Lorentz factor include the horizon shell, the surface density squared, the number of correlation patches, the particle energy, the rate of particle creation, and the transformation to the observer frame. In the end we obtain exactly the three powers which we require to push particles to the peak of the Planck spectrum in the horizon shell by the Lorentz transformation.  For the duration of the inflationary period the number density and energy density in the horizon shell are exactly at the background limit, and stay there until the refilling begins to fail.  This compensates the power-law term of the Planck spectrum of $(\hbar \omega)^{3} \frac{\hbar d \omega}{\hbar \omega}$, after taking into account the factor of $\hbar\,\omega$ the denominator. 

In the exact formulation of the background spectrum the exponent of the exponential term in the denominator has to be written as (see section II)
\be
e^{\{\sqrt{x^{2} + 2 y}\}} - 1
\ee
with $y$ time-dependent, and $x$ the normalized graviton energy, defined above in Eq. \ref{defx}.  

At the end of inflation the dimensionless parameter $x$ has the value 
\be
x \; = \; \frac{h \nu_0}{m_{Pl} c^2} \; \simeq \; 10^{-18}
\ee
The dimensionless parameter $y$ is
\be
y \;  \simeq 
\frac{c \kappa_{0}}{m_{Pl} c^2} 41 \tau_{Pl}\, . 
\ee
The parameter $2 y$ reaches the same order of magnitude as $x^2$ at the end of inflation provided that $\kappa_{0} \simeq 10^{+50} \, erg/cm$ or in terms of Planck units $\kappa_{0} \simeq \epsilon \, m_{Pl} c^2/l_{Pl}$, where $\epsilon \, \simeq \, 1$ now in dimensionless units.   When $y$ reaches the same order of magnitude as $x$, the slope of the graviton energy spectrum steepens, which effectively brings inflation to an end.  Gravitons continue to be created, but at much lower energies and in much lower numbers.  The value of the parameter $\epsilon \simeq 1$ can be identified as the point where a phase transition occurs; the Higgs boson begins to appear, and the force of gravity begins to diverge from the other fundamental forces and inflation ceases.

The model which we propose for the inflationary phase of the universe is that on the horizon shell the energy and particle density are maximal (Planck level), but particles lose energy with the Planck time as they propagate away from the horizon shell.

Going backward in time from today, we know that the dominant energy density observed today, which unlike dark energy was not created by drawing from a background field, should correspond to the maximal particle energy and particle energy density at the value of the redshift at which inflation stopped.  There are three possible energy densities today, which may relate to this epoch: a) The photon density, at around $10^{-12} \, erg \, cm^{-3}$ today, b)  the matter density, and c) a possible unknown gravitational wave energy density at a level below dark energy as we observe it.  The energy density at the end of inflation near the horizon shell is  $m_{Pl} c^{2} l_{Pl}^{-3} \, = \, 10^{115.3} erg \, cm^{-3}$ in our model.  To reach this level starting from today's dark energy  implies a redshift of $10^{-8} \,  (1 + z_{infl})^{4} \, = \, 10^{115.3}$ in units of $erg \, cm^{-3}$, which implies for radiation $(1 + z_{infl}) \, \simeq \, 10^{30}$.   Any gravitational wave background below dark energy is so highly dependent on the model, that it is hard to pin down.  In our specific model it would be the redshifted Planck spectrum, which is a very steep spectrum in energy or frequency up to an energy of $m_{Pl} c^{2} (1+z_{infl})^{-1}$. This is a frequency of approximately $10^{+10.5} \, {\rm Hz}$, at an energy density of about $10^{-12.7} \, erg \, cm^{-3}$ (using $z_{infl} \, \simeq \, 10^{32}$), corresponding to a contribution to closure density of about $10^{-5.7}$ in dimensionless units.  This is consistent with canonical models, but in our approach of course the inflationary period is completely dominated by gravitons, at maximal particle energy and maximal energy density.

This then yields a redshift since the end of inflation of about $10^{30}$, crudely consistent with other models, which suggest $10^{29}$  (e.g.  \cite{san09,dev}).  This simplification also reduces the left-over residual graviton emission spectrum.  

Using $10^{29}$ as the value of the redshift at the end of inflation, we obtain for the horizon at that redshift a distance scale of $10^{28.2 - 43.5}\, {\rm cm} = 10^{-15.3} \, {\rm cm}$ for $r$ in Eq.\ref{req}.  Since we start from a Planck length, $\sim 10^{-33} \, {\rm cm}$, this implies an increase of the radius through the inflationary period by a factor of $10^{17.7}$, corresponding to about 41 e-folds, and an increase in the number of gravitons by a factor of $10^{35.4}$.  Since this is an estimate, the  error bars are large.

This corresponds to a horizon shell Lorentz factor of $10^{17.7}$ (see above) at the end of inflation. This in turn implies that particles are created from the background with an energy at a factor below the Planck energy, that is an energy of about $10^{-1.4} \, {\rm erg} \, \simeq 10^{10.4} \, {\rm eV}$.  This is an interesting result because this is an energy which is typical of the energy of particles observed in high energy cosmic rays (e.g. \cite{sta}); we note below that this corresponds crudely to the Higgs mass.  We emphasize that this is a particle energy in an environment at maximal energy density, thus this result has no obvious relevance for our environment of today, since the energy density is lower by a factor of more than $10^{100}$.

This concludes the use of Eq.\ref{creationeq} to describe the cosmological phenomena of dark energy, black holes, and inflation.  In the following sections we analyze the predictions and consequences of the model.
\section{Predictions of the model}

Any model of physical processes is of real value only if makes predictions which are observable.   Does the model described above have any consequences which are testable by observations that can be expected to be performed within the next few years? Fortunately, the answer is yes.

In this section we discuss the various contributions to the gravitational wave background and the experimental techniques which can be used to detect those gravitational waves which constitute the effect known as `dark energy'.

There are three experimental tests which can be performed to determine the validity of these ideas: i)  pulsar timing arrays (PTAs) to detect the gravitational wave background which makes up dark energy, ii) Lunar laser ranging to detect the single solitons coming through, and iii) ultra high precision clocks to detect the noise due to the individual solitons coming through.

\subsection{Primary source of dark energy - stellar-mass black holes vs. super-massive black holes}


Since black holes of any size can stimulate graviton emission from the Planck sea, a comparison of the relative contributions to dark energy from stellar-mass black holes and super-massive black holes is in order. A useful piece of information is, whether or not the super-massive black holes have a characteristic low mass, and observations strongly suggest that they do  \cite{gre1,gre2,gre3,gre4,car}. So it is plausible to suggest that super-massive black holes may have all started with a mass near this low cut-off.  

One question relevant for the entire approach described in this paper is when the earliest super-massive black holes might have formed.  In \cite{bie} it was shown that for one specific model of dark matter star formation could be initiated very early (asee also \cite{gil07,dev10})); obviously there may be other models also allowing rather early formation of massive stars.  Since massive stars usually form in dense groups, stellar agglomeration is possible (\cite{sand,por1}).  It was shown in reference \cite{cote} that the correlations between galaxy parameters and super-massive black hole mass apparently merge into a correlation with nuclear star clusters, with a different slope 
\cite{Balcells07,Graham12,Scott12,Leigh12,Neumayer12}.  On the other hand, massive stars at near-normal chemical abundances have powerful winds limiting their mass to several hundred Solar masses (\cite{yun}). For a heavy element abundance close to zero, such winds do not exist, and the first generation of super-massive stars can form and produce the first super-massive black holes.  At about  $10^{6} \, M_{\odot}$ such stars suffer from an instability due to a combination of high radiative pressure, and subtle effects of General Relativity (\cite{app1,app2}), exploding readily.  They could then give rise to the first super-massive black holes, with masses between $10^{6}$ and $10^{7}$  $M_{\odot}$, allowing for some larger mass due to infall. Therefore, we adopted above as our reference parameters  $3 \cdot 10^{6} \, M_{\odot}$ for the black hole mass and a redshift of creation of order 50, the value for the Lorentz factor derived above results in an ``occupation number", of order unity, as required by our earlier condition.

This result implies that stellar mass black holes today violate the occupation number condition, since they drop out at a very small luminosity distance, of order $10^{-12}$ of the luminosity distance today versus redshift of order 50.  Therefore the stellar mass black holes do not contribute significantly to the steady creation of gravitons drawing the energy from the background; on the other hand, they do keep on creating new gravitons in accordance with the creation equation, just at a very low level.  However, the gravitational waves generated at their creation should produce a background at the corresponding frequency, which may be detectable with normal gravitational wave detectors; this is the long expected background.

The above analysis shows that stellar mass black holes do not make a significant contribution to dark energy. However, they do produce a gravitational wave background, and they do connect very weakly with the background Planck sea.

\subsection{Gravitational wave background: primordial background}

There is a primordial gravitational wave background at very high frequency left-over from the inflationary period (e.g. \cite{all}).  The peak frequency given by the horizon shell should be near the Planck energy divided by $10^{32}$, giving a frequency of $10^{10.5}$ Hz, corresponding to the last Planck energy gravitons produced in the horizon shell at the end of inflation.  However, the spectrum should extend steeply to lower frequencies, with the very lowest frequency being the Planck energy divided by the redshift factor as measured from the beginning of the universe of $ \sim \, 10^{47.7}$, giving a frequency of  $\sim \, 10^{-4.2}$ Hz.  For simplicity we have used a delta-function approximation for the background Planck sea of gravitons, but in reality the gravitons created from the background have a Planck spectrum.  There is an extension to even lower frequencies, again with a steep spectrum.  The energy density of this background is not far below the microwave background energy density.

\subsection{Gravitational wave background: super-massive black holes}

\begin{figure}[h]
\centering
\includegraphics[bb=0cm 0cm 21.0cm 29.7cm,viewport=2.5cm 16.5cm 16.0cm
25.7cm,clip,scale=1.2]{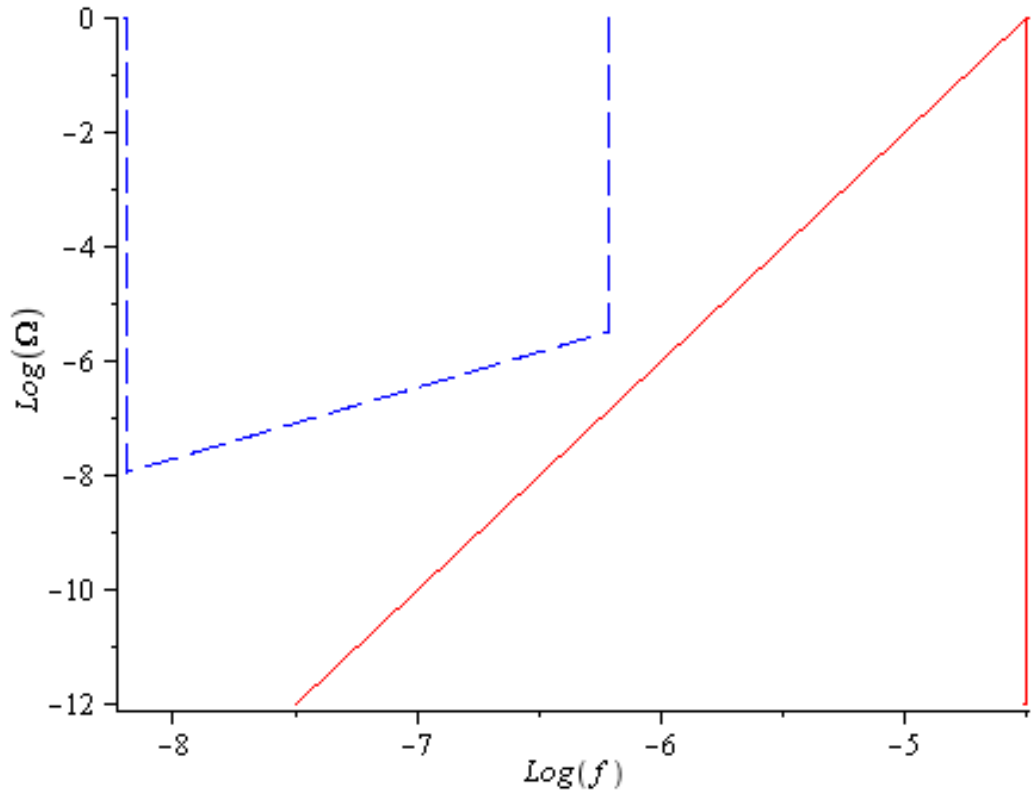}
\caption{Gravitational wave background limit from pulsar timing (dashed lines: \cite{sen1,sen2}), and our inferred gravitational wave background from stimulated emission of gravitational waves from the background Planck sea constituting dark energy, using the frequency scaling given in Eq. \ref{nuGWeq}.  We use as the ordinate the fraction of closure energy density $\Omega$, and as abscissa the frequency of the gravitational waves $f$.}
\label{DefAngles}
\end{figure}

The gravitational wave background whose existence we propose here is distinct from the primordial wave background. The gravitational waves in our model arise from the production of black holes in the early universe at a redshift of (order) 50. These black-hole-produced gravitational waves maintain a constant dark energy density by continuously interacting with the Planck sea background.  In the simplistic framework that all black holes were created at the same time, and with the same mass, this implies that the gravitational wave background peaks near 

\begin{equation}
f_{GW, max} \; = \; \frac{1}{8 \pi} \; \frac{1}{h} \, m_{Pl} c^{2} \; \frac{m_{Pl}}{M_{BH}} \, \frac{1}{1 + z_{\star}}\; \simeq \; 10^{-4.5} \, {\rm Hz} \; M_{BH, 6.5}^{-1} \, \frac{1 + 50}{1 + z_{\star}} \, ,
\label{nuGWeq}
\end{equation}

\noindent with a steep spectrum at lower frequencies, matching the Planck spectrum background, and a sharp cutoff at higher frequencies.  In Fig. \ref{DefAngles} we show a graph giving current limits on the gravitational wave background from pulsar timing (\cite{sen1,sen2}).

In this graph we note that using lower frequencies or higher initial black hole masses would approach the existing limits and thus can be ruled out today.  We propose that extending the observational sensitivity right at the point of inflection of the observational limits, near $\Omega \, \simeq 10^{-6}$ and $f \, \simeq \, 10^{-6.2} \, s^{-1}$ offers the best chance to confirm or refute the proposal made here.  A further improvement of the observational sensitivity by one or two orders of magnitude may be required to carry out this check.

Considering the many different contributions from merger events giving lower frequencies and smaller redshifts, the low frequency side of the spectrum might be a bit shallower than suggested by just the first formation of super-massive black holes.  Even matching the known dark energy contribution to the energy density of the universe, this background cannot be detected today. However the sensitivity needed to detect this background is close to the sensitivity of the planned extension of the pulsar timing arrays (e.g.  \cite{kra,liu}).   With today's limits from pulsar timing arrays it may already be possible to constrain the mass of the first generation of super-massive black holes.  We predict that this background will be detected.

\subsection{Gravitational wave background: accreting black holes}

Accreting and merging black holes emit a large fraction of the accreted mass/energy into gravitational waves, creating a gravitational wave background (e.g. \cite{wyi}).  
This background folds the temporal evolution of the growth of black holes with their mass distribution.  The contribution of such gravitational waves to the background can be estimated by the current mass density of black holes which have a typical redshift of growth.  The current mass/energy density is about $10^{-16.9} \, erg/cm^{3}$.  This value has been corrected for the redshift assumed above for the creation of the first generation of super-massive black holes, with a large error bar, as noted above.  There is an additional uncertainty due to the lack of knowledge about which process is dominant in the growth of black holes -- accretion or mergers. In this paragraph we confine our  discussion to accretion.  If the accretion process is less important for growth than mergers, then this background is weaker than that of the other processes, possibly much weaker.  In addition, since most of the growth of black holes was at redshift of approximately 2 or larger, there is a downward correction of order 10 - 20, giving an energy density of $10^{-17.9} \, erg/cm^{3}$.  This translates into a fraction of the critical density of about $10^{-9.9}$, clearly below current limits.  The proper frequencies correspond to the mass range of super-massive black holes, which range from about $10^{6} \, M_{\odot}$ to $3 \cdot 10^{9} \, M_{\odot}$.  Improvement in pulsar timing techniques may allow detection of this background some time within the next few years \cite{kra}.  In contrast to the contribution to the background from dark energy, this accretion contribution to the background is steady. The dark energy contribution is divisible into single soliton waves, or shell fronts of gravitational waves, which vary in time.

\subsection{Gravitational wave background: stellar-mass black holes}

The creation of stellar-mass black holes, which are expected to be more common in the early, low heavy-element universe \cite{mir}, also provides a gravitational wave background.  Their typical frequency is
\begin{equation}
f_{GW, stars} \; = \; \frac{1}{8 \pi} \; \frac{1}{h} \, m_{Pl} c^{2} \; \frac{m_{Pl}}{M_{BH}} \; \simeq \; 10^{+2.0} \, {\rm Hz} \; M_{BH, \odot}^{-1} \, \frac{1 + 50}{1 + z_{\star}} \, .
\end{equation}
The level of this background at the time of formation may have been an order or two orders of magnitude above that generated by the formation of super-massive black holes.  However, stellar black holes do not give rise to the shell fronts that keep up the filling condition which is necessary in order to draw energy from the background. The gravitational waves generated during the creation of stellar-mass black holes expand just the same as any relativistic gas, that is with a redshift factor of $(1+z)^{4}$ from creation. Assuming as a reference an initial redshift in the range of 50 to 80 \cite{bie}, their energy density is diluted today by about $10^{7}$.  The existing observational constraints \cite{lig} give an upper limit of about $10^{-5}$ of closure density, so at this limit the mass density of stellar black holes could exceed  that of super-massive black holes by a factor of 100, and still be allowed.  However, this demonstrates that a detection may be possible within the foreseeable future with laser interferometers.  This specific background of gravitational waves should have a time dependence just like the corresponding background from the formation or the merger of super-massive black holes.  In addition, super-massive black holes should provide a contribution to a steady background of gravitational waves due to accretion episodes.

\subsection{Lunar laser ranging}

The increase in dark energy is due to the large number of individual soliton fronts emanating from the creation and mergers of black holes in the early universe.  Each individual soliton front carries a large energy, taking into account the original rest mass energy of the black hole, and the large subsequent increase in energy due to the creation of resonant gravitons by the solitons interacting with the background Planck sea.  This energy $E_{s}$ can be estimated by
\begin{equation}
E_{s} \; = \; \frac{1}{2} \, M_{BH} \, c^{2} \, {\left(\frac{1+z_{\star}}{1+z}\right)}^{3}\, .
\end{equation}
For $M_{BH} \, = \, 10^{6.5} \, M_{\odot}$ and $z_{\star} \, = \, 50$ this energy is today ($z = 0$) $10^{65.6}$ erg.  This energy is distributed over a volume given by $4 \pi d_{L}^{2} = 10^{61.66}$ cm$^2$; the distance scale corresponding to the age of the universe is $10^{28.11}$ cm, and the corresponding volume is $10^{89.3}$ cm$^3$.  The energy density per soliton of is thus $10^{-23.7}$ erg cm$^{-3}$.  The overall change in dark energy, if it were to just relax without any re-creation, would correspond to $10^{-8} \, {\rm erg \, cm^{-3}} \, 4 \, H(z)$ or about $10^{-25.0} \, {\rm erg \, cm^{-3} \, s^{-1}}$.  With these simplistic numbers we obtain a front coming through every 20 seconds, but the transient time is of the order of 1000 seconds.  Therefore, any given moment there are roughly 50 uncorrelated shells going through.   This time between shell fronts scales linearly with $M_{BH}$ and also linearly with $(1 + z_{\star})$, so the uncertainty of this estimate of 20 seconds is probably an order of magnitude.  These soliton fronts are uncorrelated, although large scale structure should already be relevant even at $z = 50$. However it is hard to see how  the birth pangs of different black holes could influence these fronts in a statistically detectable way.  The round trip light travel time to the Moon is about 3 seconds, so a soliton passes through the Earth-Moon system roughly every ten round-trip travel times using the numerical estimates discussed above.  The best way to detect the corresponding fluctuations would be to have an electromagnetic signal travel many times from the Earth to the Moon and back to see if there are transient shifts in an interference pattern every now and then (e.g.  \cite{ste}).

\subsection{Ultra-precise timing}

In a very similar vein, ultra-precise timing experiments which test the steadiness of timing over time scales of order a few seconds to a few minutes should show these shifts as soon as the precision is high enough.  Clearly the precision to detect such a signal has to  correspond to seconds in the expansion rate of the universe, that is the precision  must be of order $10^{-17.5}$, or a few $10^{-18}$.  This precision is expected to be reached in the next generation of clocks \cite{did,chou,chou2,parthey,predehl}.

Since a single soliton coming through has at least a transition time of the inverse of the frequency given above, the order of the transit time is some fraction of an entire day. New solitons come through on the order of magnitude of every 20 seconds, so how can we expect to be able to measure this?  We note that a single soliton is a coherent superposition of approximately $10^{97}$ correlated waves (using our adopted parameters, and counting across the universe) combining to make up the soliton transition.  The different solitons coming through every 20 seconds are incoherent.  A careful error analysis might disentangle the two aspects.

\subsection{Summary of predictions}

This model makes quantitative predictions of several phenomena.  Of the predictions discussed in this section, those in Sections C and G are the most likely to be verified or refuted in the near future.  The predictions in the remaining sections are of cosmological importance, but they may require more time to test.  

\section{Consequences of the Model}

\subsection{Primordial black holes}

In the approach described here primordial black holes grow - feeding on the background - as
\be
\frac{M_{BH}}{m_{Pl}} \; = \; {\left(\frac{3 t}{ \tau_{Pl}}\right)}^{1/3}\, ,
\ee
giving them a present mass of $3\times 10^{16} \; {\rm g}$. 

In a speculative hypothesis in which all dark matter is black holes, a possibility which is not easily excluded (\cite{abram}), all primordial black holes would grow to $3\times 10^{16}$ g. Therefore in our Galaxy the total number of primordial black holes is limited by
\be
N_{PBH} \; 3\times 10^{16} \, {\rm g} \; \leq \; 10^{12} \, M_{\odot} \, \simeq 10^{45.3} \, {\rm g}\, ,
\ee
using the total mass of our Galaxy.  If we hypothesize that all primordial black holes were initially at the Planck mass - an assumption from which it is straightforward to generalize - the relevant number of primordial black holes for our Galaxy is
\be
N_{PBH} \; \leq \; 10^{29.3} \, ,
\ee
and their total mass is
\be
N_{PBH} \, m_{Pl} \; \leq \; 10^{24.3} \, {\rm g} \, .
\ee
Comparing this to the total mass of the universe we see that 
\be
\frac{N_{PBH} \, m_{Pl}}{10^{12} M_{\odot}} \; \leq \; 10^{-20.2}\, ,
\ee
and therefore
\be
\Omega_{PBH, 0} \; \leq \; 10^{-20.7}\, .
\ee
If equality in this expression were true, then all primordial black holes would account for all dark matter.  

We note here an important hidden assumption:  We implicitly assume that the primordial black holes do not get eaten by bigger black holes.  Primordial black holes may be nearly at rest with respect to the microwave background system, and so may easily be eaten by the largest of the super-massive black holes.  If that is an important process due to the evolution of large scale structure, then these limits derived here do not apply.

There is a secondary difficulty in that under this hypothesis dark matter would not be constant, but would vary with cosmic epoch.  Using the simple power-law approximation the mass dependence of the primordial black holes as a function of redshift would be
\be
\frac{M_{BH}}{m_{Pl}} \; = \; {\left(\frac{3 \tau_{H}}{ \tau_{Pl}}\right)}^{1/3} \, \frac{1}{(1 + z)^{1/2}}\, ,
\ee
which can probably be excluded already with existing data as the possible redshift dependence of dark matter.  If this can be reliably proven, then the upper limit given above is firm.  This limit is much stronger than discussed in \cite{abram}.

We emphasize again, that black holes of low mass do not fulfill the occupation number condition on spawning an active shell front.  However, this condition just states according to the creation equation, that creation of new gravitons is strong, and perhaps self-sustaining.  The creation equation clearly shows, as already emphasized above, that creation does continue, but at an extremely low level.  Therefore, new gravitons are created all the time, but at a level energetically completely irrelevant compared to dark energy driven by the shell fronts from super-massive black holes.

\subsection{Lorentz Invariance}

It is clear from section IV.A that Lorentz Invariance is violated in our model in the limits of very high energy, a very large number of correlated bosons, very small length scales, and very small time scales.  For instance, the minimum scale both in space and in time cannot be under-cut in any frame; similarly, the maximum energy cannot be exceeded in any frame.  This is consistent with the Horava - Lifschitz theory of gravity in which Lorentz invariance is violated at very high energies or equivalently at very small distance scales \cite{hor,vis}. Even singularities cannot be pin-pointed with a precision better than these limits. This has led
us to shells instead of horizons, and to a finite Lorentz factor for the shell front.  We propose to deal with the further consequences elsewhere.

\subsection{Astronomical predictions}

In the scenario proposed here the main population of super-massive black holes has to be formed reasonably early to allow dark energy to be started.  The efficiency of their contribution strongly decreases with redshift, and so dark energy is predicted to be very nearly constant to a fairly high redshift range, consistent with very recent work, \cite{Reid}.

Our approach strongly suggests that super-massive black holes grow mostly by merging, which should be testable with various observational strategies.

Another consequence of our model clearly is, that reionization is extremely clumpy, and starts even before the first super-massive black holes, which in our model are formed out of the agglomeration of stars. Our model thus requires the early  formation and demise of stars, and predicts the detection of signatures such as heavy elements, dust, magnetic fields, and cosmic rays for such stellar events early in the history of the universe.  Secondary consequences  derive from cosmic ray interaction, such as high energy neutrinos; neutrinos should be strongest from the explosion of the first massive stars with sufficient heavy elements to form stellar winds to explode into (see, e.g., \cite{CRII}).

\subsection{Arrow of time}

In our model time is driven by the energy transfer from the Planck sea to our universe.  Thus the arrow of time is uni-directional, since in our model energy flows in only one direction - from the background into our universe.  An immediate consistency check of our model is the question of whether or not it describes a universe which is connected to the background Planck sea at all times.

The horizon shells dip below criticality at the end of inflation, but at a much reduced level they still create new gravitons, allowing a weak connection to the background throughout the transition phase from the cessation of inflation to the first creation of super-massive black holes.

The shell fronts produced by the formation and mergers of the super-massive black holes come through at an estimated rate of one every twenty seconds, and any specific transition lasts a fraction of a day.  So at any given moment we experience of order one thousand shell fronts, all connecting us to the background Planck sea.

The shell fronts produced by stellar black holes or Planck particles do not give the full occupation number, and so their creation of new gravitons is enormously reduced, by many powers of ten.  However, this creation is not zero.  The number of stellar black holes is of order possibly $10^{9}$ higher than the number of super-massive black holes, and their formation probably peaked relatively early in the universe, when the heavy element abundance was significantly lower (e.g. \cite{mir}).  Therefore their repetition time is very much shorter (the time from one to the next), as is their transition time (the step-length, which is essentially the inverse gravitational wave frequency).  So over a very wide range of time-scales there is always a connection to the background, albeit extremely weak.  So the arrow of time is always assured.

In the far distant future, when the shell fronts of the super-massive black holes become sub-critical, causality will always be assured by the decreasing interaction with the background.  The universe will no longer be driven by dark energy, as it will turn into a relativistic gas, expanding essentially adiabatically.  The universe will end altogether, when this tenuous connection to the background lapses, since causality is no longer possible.

\subsection{Signal speeds and mass}

The Lorentz factor for the effective propagation of a signal derived above should apply to the propagation of any signal carried by mass-less boson particles, or waves travelling with the speed of light even at distances as small as sub-atomic scales.  In our model such signals never travel at exactly the speed of light, but always just very slightly below. The limit is given by the number of coherent waves (see Eq. \ref{coherenceeq})
\begin{equation}
\Gamma_{r} \; = \; {\left( \frac{d}{8 c h} \, {\left( \frac{E_{1} \, E_{\Sigma}}{\pi}  \right)}^{1/2} \right)}^{1/2} \, ,
\end{equation}
where $E_{1}$ is the energy of a single boson, $E_{\Sigma}$ is the total energy of all coherent bosons, and $d$ is the distance travelled.  The smallest possible limit is obviously the Planck length, and at that limit the Lorentz factor becomes
\begin{equation}
\Gamma_{r} \; = \; {\left( \frac{d}{4 \, l_{Pl}} \,  \right)}^{1/2}\, .
\end{equation}
At the smallest distances sometimes considered in particle physics, of order  $\sim \, 10^{-18} \, {\rm cm}$, this Lorentz factor would still be $\Gamma_{r} \simeq 10^{+7.5}$, but would require, independent of the length $d$, that
\begin{equation}
E_{1} \, E_{\Sigma} \, = 2 {\left(2 \pi \right)}^{3} \, {\left(m_{Pl} \, c^{2}\right)}^{2} \; \simeq \; 10^{35.3} \, {\rm erg^{2}} .
\label{EEsum}
\end{equation}
It is straight-forward to verify that shell fronts obey this condition.

However, there is a limit to this kind of reasoning, since the smallest possible energy unit in the universe is the energy of a gravitational wave spanning the entire universe. This energy is
\be
E_{1} \; = \; \frac{c h}{\lambda_{univ}} \, \simeq \, 10^{-43.9} \, {\rm erg}\, .
\ee
If the radius of the universe is taken to be a half wavelength, then the length scale is twice as large and the energy is half as large.  This energy is the equivalent energy obtained by using the Gibbons-Hawking temperature (\cite{gibhaw77,gibhaw79}).  At high redshift this scales as $(1+z)^{3/2}$, but with a factor 1.3 due to the subtle effect of dark energy. At the end of inflation a phase transition occurs.  Requiring that the phase transition parameter $\epsilon = 1$ in natural units, the energy $E_1$ is $10^{-0.8} \, {\rm erg}$ which implies that ($z_{infl} \, \simeq \, 10^{29}$)
\be
E_{1,infl} \simeq \, 100 \, {\rm GeV} \; {\left(\frac{z_{infl}}{10^{29}}\right)}^{3/2}\, ,
\ee
as the energy where mass first begins to manifest itself in the evolution of the universe.  When the first rest mass appears, the scaling of energies with look-back time fails and inflation stops.   This value is obviously subject to the uncertainty of when inflation really ends.  This can also be written (see Eq. \ref{inflation}) to within a factor of order unity as
\be
E_{1,infl} \simeq \, m_{Pl} \, c^2 e^{-t_{infl}/\tau_{Pl}} \, .
\ee
Integrating over all gravitons created during inflation by number and energy shows that Eq. \ref{EEsum} is obeyed right through the inflationary period, with $E_1$ always the lowest energy in the system.

\begin{figure}[h]
\centering
\includegraphics[bb=0cm 0cm 25.0cm 29.7cm,viewport=5.0cm 3.0cm 26.0cm
22.7cm,scale=0.38]{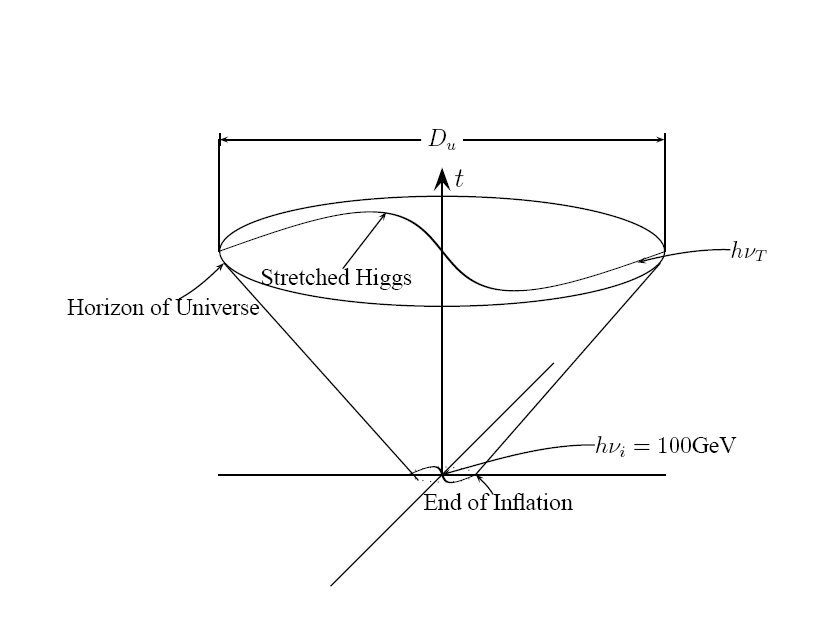}
\caption{The lowest energy graviton existing today is the redshifted lowest energy graviton at the end of inflation.  At the end of inflation the energy of this graviton was $\simeq 100$ GeV.}
\label{higgs}
\end{figure}

This is a crude estimate of the individual mass of a boson at the end of inflation, where we are able to reach the Planck uncertainty, imposing a behavior that first gives the outside appearance of mass.  This energy is of the same order of magnitude as that of the Higgs boson (see Fig.\ref{higgs}).  

In a future investigation we will carry out a more refined
mathematical calculation to determine whether or not there is a
connection between the signal mass at the end of inflation and that of
the Higgs boson.

This is the signature of a finite rest-mass. The fact that wave pulses travel at very slightly subluminal velocities is due, in the large number limit, to the uncertainties in space and time at the Planck level.

\subsection{Consistency}

The model of the generation of gravitons by shells (shell fronts, Planck shells, or horizon shells) is not a fully developed theory.   What we have shown is that an approach exists, which given its assumptions can provide a common basis for the understanding of such diverse phenomena as black holes, dark energy, inflation, rest mass, and the arrow of time. Clearly this is merely a first step in the development of a complete theory.

A fully developed theory would start with some prescription for the production of various black holes and their interactions including mergers. A complete theory could describe their formation, growth and merger history (e.g.  \cite{sand,abel,berti,bello,ger,por1,por2,por3,por4,vol1,vol2,vol3,li,fan}), and at the same time describe the evolution of their shell fronts, which contribute to dark energy.  Over a certain mass range black holes drop out of creating dark energy, since they can no longer provide the critical surface density on their shell front. The description of all these processes can be given in terms of functions of the redshift $z$.  Dark energy enters the integrals for luminosity distance and Hubble parameter evolution, turning the creation equation into a set of coupled non-linear differential and integral equations.  The overall evolution of the universe is then governed by the creation equation integrated over all of the black holes, over their creation and over their merger history.

In the three applications of the creation equation we used gravitons pulled from the background usually below the maximum energy in the background, so that with the proper Lorentz factor they are at the Planck peak on the shell.  The actual number pulled is far below the density in the background for the two applications of the shell front and the black holes. But only for the horizon shell, so for the inflationary phase, the density of gravitons pulled in is so high, that modifications of the background spectrum due to the appearance of separate forces of nature do become important.  Only in that case do we reach the limits of the phase space density in the background itself.

The model we propose in this paper could be disproved in several ways:  a)  if the gravitational wave background corresponding to dark energy is not detected with the appropriate strength and spectrum; b)  if the formation of the super-massive black holes is near redshift 10 or below rather than near 30 - 50; or c) if the ultra-precise timing and lunar laser ranging does not detect the individual soliton fronts of gravitational waves.  If, however, these predictions were 
confirmed, new paths of investigation would open up for research on dark 
energy, black holes, inflation, Lorentz transformations, the arrow of 
time, and the characteristic energy at which rest mass appears.

\subsection{Summary}

We have introduced the notion of stimulated emission of gravitons on a critical surface, drawing energy from a background field. The phase space density used in our model is derived in close analogy to the holographic principle.  The equation derived in Section III for the stimulated emission of gravitons from a background is the first new element in our approach, and the use the equation for the very large Lorentz factor derived from Planck uncertainties is the second new element. This allows the interpretation of dark energy, black holes, the inflationary period of the universe, the arrow of time and the characteristic energy at which rest mass first manifests itself to be put on a common basis.  The other concepts used in our model are variants of commonly discussed ideas in gravity theory.

The next step is to develop a more formal mathematical theory to self-consistently derive the creation equation.  Part of this formal development will include a proper treatment of Lorentz transformations.  An additional step is to include the spin of the black holes, and   another is to explore some of the consequences of our model in more detail, such as the signal speed limit; the signal speed limit directly relates to the characteristic energy at which rest mass first appears.  Most importantly, we will develop more specific proposals for experimental tests of this general approach.

\acknowledgments

Discussions with Lou Clavelli greatly contributed to the development of the paper; discussions with Lauren\c{t}iu Caramete (Bucharest, Romania), Roberto Casadio (Bologna, Italy), Marco Cavaglia (Oxford, MS, USA), Laszlo Gergely (Szeged, Hungary), Shaoqi Hou (Tuscaloosa, AL), Pankaj Joshi (Mumbai, India), Gopal-Krishna (Pune, India), Octavian Micu (Bucharest, Romania), Piero Nicolini (Frankfurt, Germany), Norma Sanchez (Paris, France), Joe Silk (Oxford, United Kingdom), C. Sivaram (Bangalore, India), Allen Stern (Tuscaloosa, AL), Dijan Stoikovic (Buffalo, NY, USA),  and Hector de Vega (Paris, France) are gratefully acknowledged.  Helpful comments on earlier versions of the manuscript were received from Roberto Casadio, Marco Cavaglia, Laszlo Gergely, Alister Graham, Pankaj Joshi, Octavian Micu, Piero Nicolini, and Dijan Stoikovic.

This research was supported in part by the DOE under grant DE-FG02-10ER41714.  

\section{Appendices}

\subsection{ Graviton distribution function}

The graviton distribution function on the weak-gravity brane (our brane) for a Planck surface interacting with a background field can be determined from the Boltzmann equation for coherent production of gravitons in a Friedman-Robertson-Walker (FRW) universe \cite{wei}, see \cite{bedo90}.  The distribution function ${\mathcal N}(k,t)$ satisfies the equation ($\hbar\,=\,c\,=\,1$ for the derivation of the differential equation satisfied by ${\mathcal  {N}}(k,t)$)
\be
\left(\frac{\partial}{\partial t}-\frac{\dot{R}(t)}{R(t)}\, k\,\frac{\partial}{\partial k}\right)\,{\mathcal  {N}}(k,t) &=& \frac{1}{k}\int\,\frac{d^3\, k'}{(2\pi)^3\,2\, k'}\int\,\frac{d^3\, p}{(2\pi)^3\,2\, E(p)}\int\,\frac{d^3\,p'}{(2\pi)^3\,2\, E(p')}\int\,\frac{d^3\, k''}{(2\pi)^3\,2\, k''}\nonumber\\
&&|M|^2\,2\,\pi^4\delta^4(k+p-k'-k''-p')\,2\,\pi^3\,\delta^3(\vec{k'}-\vec{k''})\nonumber\\
&&\left({\mathcal  {N}}(k',t)\,{\mathcal  {N}}(k'',t)\, g_{b}(p',t)\,(1+{\mathcal  {N}}(k,t)) \right.\nonumber\\
&&- \left. {\mathcal {N}}(k,t)\, g_{b}(p,t)\,(1+{\mathcal {N}}(k',t))\, (1+{\mathcal {N}}(k'',t))\right)
\ee
The $\delta$-function, $\delta^3(\vec{k'}-\vec{k''})$, has been inserted to impose coherence of the outgoing gravitons.  $|M|^2$ is the matrix element squared for the quadrupole emission of a graviton of energy $k''$.  $g_b(p,t)$ is the occupation number distribution of the background particle sea.  $R(t)$ is the scale factor for an expanding universe.

The 4-dimensional $\delta$-function can written as
\be 
\delta^4(k+p-k'-k''-p') = \delta(k_0 +E_p - k_0'-k_0''-E_p')\,\delta^3(\vec{k}+\vec{p}-\vec{k'}-\vec{k''}-\vec{p'})
\ee
 Imposing the coherence condition and assuming that the momentum transfer $\vec{k}-\vec{k'}$ is small, the $\delta^3$-term can be expanded in a series
 \be 
 \delta^3(\vec{k}+\vec{p}-2\,\vec{k'}-\vec{p'}) &=& \delta^3(\vec{p}-\vec{p'}-\vec{k'})\,+\,(\vec{k}-\vec{k'})  \frac{\partial}{\partial q'}\delta^3(\vec{p}-\vec{p'}-\vec{k'})\nonumber\\&&+\,\frac{1}{2}(\Delta\vec{k}\cdot\frac{\partial}{\partial q'}\delta^3(\vec{p}-\vec{p'}-\vec{k'})+\cdots 
 \ee
 where $\vec{q} = \vec{p'}+\vec{k'}$.  If the background particles are assumed to have structure, to have excited states and to obey the relation $\vec{p}\,\simeq\,\vec{p'}$, then first term in the expansion of $\delta^4(k+p-k'-k''-p')$ can be factored as
 \be 
 \delta^4(k+p-k'-k''-p') = \delta^3(\vec{p}-\vec{p'}-\vec{k'})\,\delta(k_0-k_0')
 \ee
 where the coherence condition $\vec{k'} = \vec{k''}$ has been inserted.  The second term in the expansion vanishes when it is integrated over all angles.

The Boltzmann equation for $\mathcal{N}(k,t)$ to lowest order in the expansion of the 4-dimensional $\delta$-function is
\be 
\left(\frac{\partial}{\partial t}-\frac{\dot{R}(t)}{R(t)}\, k\,\frac{\partial}{\partial k}\right)\,{\mathcal  {N}}(k,t) \simeq \frac{-\kappa}{k}\,{\mathcal{N}}(k,t)\,({\mathcal{N}}(k,t)+1) \, ,
\ee 
where the factor $\kappa$ is given by
\be 
\kappa = \frac{1}{2\pi R(t)}\int\frac{d^3p}{(2\,\pi)^3}\frac{|M|^2}{E(p)^2} \; = \; \frac{\kappa_{0}}{R(t)} 
\ee 
and $|M|^2$, as stated above, is the matrix element squared for quadrupole emission of a graviton wave
from the background.

The product ${\mathcal {N}}(k,t)\,({\mathcal {N}}(k,t)+1)$ can be expressed as
\be 
-{\mathcal {N}}(k,t)\,({\mathcal {N}}(k,t)+1) = \frac{k_B}{c}\,T_g\,\frac{\partial}{\partial k}{\mathcal {N}}(k,t)\, , 
\ee
where $T_g = T_{g0}/R(t)$ and $k$ is the magnitude of the graviton momentum.  After making the replacement $k = \tilde{k}/R(t)$, the equation for ${\mathcal {N}}(\tilde{k},t)$ becomes
\be 
\frac{\partial}{\partial t}{\mathcal {N}}(\tilde{k},t) = \,\kappa_{0} \frac{k_B\,T_{g0}}{ c}\frac{1}{\tilde{k}}\frac{\partial}{\partial\tilde{k}}{\mathcal {N}}(\tilde{k},t)\, ,
\ee
or in terms of the frequency of the wave at emission $\nu_0$
\be 
\frac{\partial}{\partial t}{\mathcal {N}}(\nu_0,t) = \,\kappa_{0} \frac{k_B\,T_{g0}}{h^2 c}\frac{c}{\nu_0}\frac{\partial}{\partial\nu_0}{\mathcal {N}}(\nu_0,t)\, ,
\ee
\noindent The solution of this equation can be written in terms of dimensionless parameters $x$ and $y$, where
\be 
x = \frac{h\,\nu_0}{k_B\,T_0}
\ee
\noindent and
\be 
y = \int_0^t\,\frac{\kappa_{0}\,c}{k_B\,T_{0}}\frac{T_{g0}}{T_0}\, dt' 
\label{defy}
\ee
\noindent as 
\begin{equation}
{\mathcal N}(x,y)\,=\,\frac{1}{\pi^2} \, \frac{x^{3}}{e^X - 1} 
\end{equation}
with 
\begin{equation}
X \, = \, \sqrt{\{x^2 + 2 y\}} \; .
\end{equation}
When $y$ reaches the same order of magnitude as $x^2$, the number density ${\mathcal{N}}(x,y)$ begins to deviate from the Bose-Einstein distribution.  This corresponds to the appearance of the Higgs boson.

\subsection{ The Kompaneets equation}

For photons, also bosons, the basic equation is the Boltzmann equation describing the scattering of photons on electrons, and includes the differential cross-section, an integral over directions, the distribution function of the electrons, and the phase space distribution of the photons, always including stimulated emission which implies the use of the square of phase space density, see above, and also, e.g. \cite{ryb}.  In this application the equation was first derived by Kompaneets.

The relativistic Boltzmann equation describing stimulated emission has already been given above.  The non-relativistic form for stimulated emission for photons scattering from a distribution of electrons
\be
p + \omega <=> p_1 + \omega_1
\ee
is (see, e.g. \cite{ryb})
\be
\frac{\partial}{\partial t} \, n(\omega) \; = \; c \int d^3 p \, \int \frac{d \sigma}{d \Omega} \, d \Omega \left( f_e({\bf p_1}) \, n(\omega_1) \{1 + n(\omega)\} -
f_e({\bf p})  \, n(\omega) \{1 + n(\omega_1)\} \right)
\ee
where $f_e(p)$ is the distribution of electrons, $n(\omega)$ is the distribution function of photons, $\sigma$ is the cross-section, and the photon density squared term accounts for stimulated emission.

Simplifying this Boltzmann equation to the limit that the energy change is minimal, that the cross-section is isotropic, the phase space density $n(\omega)$ is very large, and the distribution functions are very peaked, we obtain
\be
\frac{\partial}{\partial t} \, n(\omega) \; = \; \sigma \, \frac{d f_e}{d p} \, \Delta \{p c\} \, n^2(\omega) 
\ee
neglecting lower terms, which do not scale with $n^2(\omega)$.  Here $\Delta \{p c\}$ is the energy being exchanged.  We will use this expression for insight as to how an equation may be structured to describe stimulated emission of gravitons from a background.   Of course this is just a didactic step, since the production of gravitons in our world by stimulated emission from the 
background is clearly relativistic.  What we want to describe is the total energy injection by stimulated emission, so we have to modify this expression to account for the energy emitted into our world, and also to account for the total phase space affected in a given time interval.

\subsection{  The end of dark energy}

Dark energy runs out of strong creation when the background phase space density becomes too low at resonance to supply enough gravitons to keep dark energy production going.  The parameter $y$ as defined in Eq. \ref{defy} reaches the same order of magnitude as $x^2$ when
\begin{equation}
y \; \simeq \, \frac{x^2}{2}\, .
\end{equation}
The future scale factor of the universe is
\be
R(t) \; = \; R_0 \, e^{H_0 t}
\ee
The parameter $y$ reaches the scale of $x^2$ at time $t_{DE}$, which
scales as 
\be 
t_{DE} \; = \; \frac{1}{H_0} \, \{-\ln {(1+z)} \} .  
\ee
and is determined from the relation
\be
x^2 \; \simeq \, 10^{-92} (1+z_{DE})^2 \, \simeq \, \frac{\kappa_{0} c T_{g 0}}{k_B T_0^2} \, t_{DE}
\ee
or in Planck units
\be 
x^2 \simeq \epsilon\,\frac{m_{pl}\,c^3}{l_{pl}}\,\frac{T_{g0}}{k_B\,T_0^2}
\ee
Solving for $z_{DE}$ implies that $\epsilon $, the dimensionless value of $\kappa_0$,  is of order $(\tau_{Pl} H_0)^2$.  A given value for $\epsilon$ determines the time at which dark energy will cease to be renewed.  Thereafter the universe will expand adiabatically.

\subsection{ The cosmological equations}

The effect of dark energy on the metric tensor elements which describe the evolution of the universe can be determined by calculating the energy-momentum tensor for a fluid whose entropy continually increases.
The metric for our universe is taken to be of the standard form as shown in the invariant line element 
\be 
ds^2 = -c^2\,dt^2 + a(t)^2\,\left(dx^2 + dy^2 + dz^2\right) 
\ee
A Lagrangian density of the form \cite{ray}
\be 
{\mathcal{L_{PF}}} = -\frac{1}{c}\,\sqrt{-g}\,(\rho + \rho_{DE})\,(c^2 + {\mathcal{E}}) 
\ee
where ${\mathcal{E}}$ is the rest, internal specific energy of the fluid, leads to the desired form of the energy-momentum tensor, provided that,
\be 
\frac{d{\mathcal{E}}}{d\rho} = \frac{P}{\rho^2}
\ee
and
\be 
\frac{d{\mathcal{E}}}{d\rho_{DE}} = -\frac{P_{DE}}{\rho^2}
\ee

The total energy momentum tensor is assumed to be the sum of two contributions 
\be 
T^{\mu\nu} = T^{(F)\mu\nu}\,+\,T^{(DE)\mu\nu}
\ee
where $T^{(F)\mu\nu}$ and $T^{(DE),\mu\nu}$ are the energy-momentum tensors for a fluid in our universe and for the dark energy in our world, extracted from the background Planck sea, respectively.  Assuming that the dark energy components, $T^{(DE)\mu\nu}$, are functions of $t$ only, $T^{\mu\nu}$ can be written as
\hspace*{-2.5cm}{\be
 T^{\mu\nu}= 
 \left[ \begin {array}{cccc} \rho
 \left( t,x,y,z \right) +{\it T_{DE}^{00}} \left( t \right) &0&0&0
\\ \noalign{\medskip}0&{\frac {P \left( x,y,z \right) }{ \left( a
 \left( t \right)  \right) ^{2}}}+{\it T_{DE}^{11}} \left( t \right) &0&0
\\ \noalign{\medskip}0&0&{\frac {P \left( x,y,z \right) }{ \left( a
 \left( t \right)  \right) ^{2}}}+{\it T_{DE}^{22}} \left( t \right) &0
\\ \noalign{\medskip}0&0&0&{\frac {P \left( x,y,z \right) }{ \left( a
 \left( t \right)  \right) ^{2}}}+{\it T_{DE}^{33}} \left( t \right) 
\end {array} \right] \; . \nonumber \\
\ee}
The conservation of energy relations are
\be
 T^{\mu\nu}_{\: ;\,\mu}= 0 
\ee
{\small\hspace*{-1.5cm}{\be 
T^{\mu\,0}_{\: ;\mu}=&&\frac{ \left( {\frac {\partial }{
\partial t}}\rho \left( t,x,y,z \right)  \right) a \left( t \right) +
 \left( {\frac {d}{dt}}{\it T_{DE}^{00}} \left( t \right)  \right) a \left( t
 \right)}{a(t)}
 +\frac{3\, \left( {\frac {d}{dt}}a \left( t \right)  \right) p
 \left( x,y,z \right) + \left( {\frac {d}{dt}}a \left( t \right) 
 \right) {\it T_{DE}^{11}} \left( t \right)  \left( a \left( t \right) 
 \right) ^{2}}{a(t)}\nonumber\\
 +&&\frac{3\, \left( {\frac {d}{dt}}a \left( t \right)  \right) 
\rho \left( t,x,y,z \right) +3\, \left( {\frac {d}{dt}}a \left( t
 \right)  \right) {\it T_{DE}^{00}} \left( t \right) + \left( {\frac {d}{dt}}
a \left( t \right)  \right)}{a(t)} \nonumber\\
&&
+\frac{ {\it T_{DE}^{22}} \left( t \right)  \left( a
 \left( t \right)  \right) ^{2}+ \left( {\frac {d}{dt}}a \left( t
 \right)  \right) {\it T_{DE}^{33}} \left( t \right)  \left( a \left( t
 \right)  \right) ^{2}}{a \left( t \right) }
 \ee}
 \be
T^{\mu\,1}_{\:;\mu}={\frac {{\frac {
\partial }{\partial x}}P \left( x,y,z \right) }{ \left( a \left( t
 \right)  \right) ^{2}}}\nonumber\\
 T^{\mu\,2}_{\:;\mu}={\frac {{\frac {\partial }{\partial y}}P
 \left( x,y,z \right) }{ \left( a \left( t \right)  \right) ^{2}}}\nonumber\\
T^{\mu\,3}_{\:;\mu}= {\frac {{\frac {\partial }{\partial z}}P \left( x,y,z \right) }{
 \left( a \left( t \right)  \right) ^{2}}}
  \ee}
The last three equations can be written as
\be 
\nabla{P} = 0 
\ee
If the dark energy is considered to be a kind of fluid, then $T^{\mu\nu}$ has the form
{\small\hspace*{-1.5cm}
{\be
T^{\mu\nu}= 
\left[ \begin {array}{cccc} \rho
 \left( t,x,y,z \right) +{\it g(t) \rho_{DE}} \left( t \right) &0&0&0
\\ \noalign{\medskip}0&{\frac {P \left( x,y,z \right) }{ \left( a
 \left( t \right)  \right) ^{2}}}+{\it f(t) P_{DE}} \left( t \right) &0&0
\\ \noalign{\medskip}0&0&{\frac {P \left( x,y,z \right) }{ \left( a
 \left( t \right)  \right) ^{2}}}+{\it f(t) P_{DE}} \left( t \right) &0
\\ \noalign{\medskip}0&0&0&{\frac {P \left( x,y,z \right) }{ \left( a
 \left( t \right)  \right) ^{2}}}+{\it f(t) P_{DE}} \left( t \right) 
\end {array} \right] \, , \nonumber \\
\ee}}
and the covariant divergence equations become
\be 
T^{\mu\,0}_{\: ;\mu} &=&\frac { \left( {\frac {\partial }{
\partial t}}\rho \left( t,x,y,z \right)  \right) a \left( t \right) +
 \left( {\frac {d}{dt}}{\it g(t) \rho_{DE}} \left( t \right)  \right) a
 \left( t \right) +3\,P \left( x,y,z \right) \frac{d}{d t}a}{a(t)} \nonumber \\ && 
 \frac{   +3\, \left( {\frac {d}{dt}}a \left( t \right)  \right) {\it 
f(t) P_{DE}} \left( t \right)  \left( a \left( t \right)  \right) ^{2}}{a(t)}\nonumber\\
&&\frac{+3\,
 \left( {\frac {d}{dt}}a \left( t \right)  \right) \rho \left( t,x,y,z
 \right) +3\, \left( {\frac {d}{dt}}a \left( t \right)  \right) {\it 
g(t) \rho_{DE}} \left( t \right) }{a \left( t \right) }
\ee
Conservation of energy requires $T^{\mu\,0}_{\: ;\mu} \; = \; 0$.  In order for the numerical factor and the powers of the scale parameter $a(t)$ to match we have to identify the central terms 
\be
\frac{d}{d t} \left( g(t) \rho_{DE} \right) + \frac{   3\, \left( {\frac {d}{dt}}a \left( t \right)  \right) {\it 
f(t) P_{DE}} \left( t \right)  \left( a \left( t \right)  \right) ^{2}}{a(t)} + \frac{3\, \left( {\frac {d}{dt}}a \left( t \right)  \right) {\it 
g(t) \rho_{DE}} \left( t \right) }{a \left( t \right) }\; = \; 0
\ee
or 
\be
\frac{d}{d t} \left( g(t) \rho_{DE} \right) + 3 \left(\frac{\dot{a}}{a}\right) \, \left( f(t) \, a^{2} \, P_{DE} \, + \, g(t) \, \rho_{DE}\right) \; = \; 0
\ee

The source term  in the conservation equation is $3 \rho_{DE} H(z)$.  This term can also be obtained by first combining Eqs.\ref{ngw} and \ref{egw} with the density, $n_{BH,0}$, of black holes to obtain
\be 
\rho_{DE} = \frac{1}{2}\,M_{BH}\,c^2\,n_{BH,0}
\ee
and then using \ref{eq1} to obtain the rate of dark energy production
\be 
\Sigma_{DE} = \frac{3}{2}\,M_{BH}\,c^2\,n_{BH,0}\,H(z)\,=\, 3\,\rho_{DE}\,H(z)
\ee
The expressions for $g(t)$ and $f(t)$ are not fixed by either Einstein's equations nor the continuity equation.  Therefore we set $g(t)\,=\,1$ and determine a reasonable form for $f(t)$ by the following argument.  Gravitational systems may have a negative specific heat \cite{spitzer69}, and examples are Schwarzschild black holes, and self-gravitating globular clusters of stars.  If BHs are in a non-flat geometry such as de Sitter or anti-de Sitter (cosmological constant different from zero), then there are values for the angular momentum (Kerr BH) and charge (Reissner-Nordstr{\"o}m BH) for which the specific heat is negative, but these are just examples.  This is exactly the situation we are proposing.  So we suggest that $f(t) \, = \, - 1/a^2$, or equivalently that $P_{DE} \, = \, - \rho_{DE}\,c^2$.  This solves, as usual, the requirement that normal matter runs as $(1+z)^3$, and that the conservation of energy is obeyed.

The Einstein equations are automatically satisfied, since we use the normal energy-momentum tensor.  With these choices for $g(t)$ and $f(t)$ the equation satisfied by $a(t)$ can now be determined.  The Einstein equations can be written as 
\be 
R_{\mu\nu} = -8 \pi G_N \left(T_{\mu\nu} - \frac{1}{2}g_{\mu\nu}\,T\right) 
\ee
where $T = T^{\lambda}_{\:\lambda}$.  The $t,t$ component of the Ricci tensor is
\be 
R_{tt} = 3\,{\frac {{\frac {d^{2}}{d{t}^{2}}}a \left( t \right) }{a \left( t
 \right) }}
\ee
and the spatial components of the Ricci tensor all have the form
\be 
R_{ii} = -2\, \left( {\frac {d}{dt}}a \left( t \right)  \right) ^{2}-a \left( t
 \right) {\frac {d^{2}}{d{t}^{2}}}a \left( t \right)\, \: i\,=\, 1,2,3\, . 
\ee
These two equations and the expressions for the $T_{\mu\nu}$ can be used to eliminate the $\ddot{a(t)}$ term, giving 
\be 
\left(\frac{da(t)}{dt}\right)^2 = \frac{8 \pi G_N a(t)^2}{3}\left(\rho(t)+ g(t) \rho_{DE}(t)\right)
\ee
This is expression for $(da / dt)^2$ is of the standard form, but with $\rho(t)$ replaced by $\rho(t) + g(t) \, \rho_{DE}(t)$; as shown above the most likely solution is that $g(t) \, = \, 1$.   The only modification which we have made is to identify the dark energy instead of subsuming it into the density as one term.

This completes this argument; our proposed theory for the origin of dark energy satisfies the energy conservation equation including the proper source term, and the Einstein equations.

\subsection{ Graviton emission from black holes}

Here we derive the steady graviton emission from black holes in a manner analogous to Parikh \& Wilczek (\cite{parikh}) in an expanding universe.

Defining
\be 
\rho = r\,e^{\beta(t)/2}\, , \:\: \mbox{and}\: m = \mu(0)\, =\, \mu(t)\,e^{\beta(t)/2}\, ; \:e^{\beta(t)/2}\, =\, a(t)\, ,
\label{rm}
\ee
Eq.(\ref{met}) can be written as
\be 
ds^2 = \frac{-(1-\frac{m}{2\,\rho})^2}{(1+\frac{m}{2\,\rho})^2}\,dt^2\,+\,\left(1+\frac{m}{2\,\rho}\right)^4\left[ d\rho^2\,+\,\rho^2\,d\Omega_2\right]
\ee

The emission of a graviton of energy $E_g = \hbar\,\omega$ lowers the mass of the black hole by the amount $E_g$ to $m-\omega$ ($\hbar\,=\, c\,=\, 1$).  So the metric after emission is 
\be 
ds^2 = \frac{-(1-\frac{m-\omega}{2\,\rho})^2}{(1+\frac{m-\omega}{2\,\rho})^2}\,dt^2\,+\,\left(1+\frac{m-\omega}{2\,\rho}\right)^4\left[ d\rho^2\,+\,\rho^2\,d\Omega_2\right] \, .
\ee 

The action for an outgoing S-wave particle is
\be 
S &=& \int_{\rho_{in}}^{\rho_{out}}\, p_\rho\,d\rho \nonumber\\
&=& \int_{\rho_{in}}^{\rho_{out}}\,\int_{m(t)}^{m(t)-\omega(t)}\frac{dH}{\dot{\rho}}\,d\rho
\label{action}
\ee
where $H\,=\,m(t)-\omega(t)\, , \: \rho_{in}\,=\, a(t)\,m\, , \: \rho_{out}\,=\,m(t)-\omega(t)$.  The null radial $\rho$ geodesics  yield
\be 
\dfrac{d\rho}{dt} = \frac{1-\frac{(m(t)-\omega(t))}{2\,\rho}}{\left(1+\frac{(m(t)-\omega(t))}{2\,\rho}\right)^3}\, .
\ee
Inserting this expression for the $\dot{\rho}$ term, the action becomes
\be 
S = -\int_{\rho_{in}}^{\rho_{out}}\,\int_0^{\omega(t)}\frac{{\left(1+\frac{(m(t)-\omega'(t))}{2\,\rho}\right)^3}}{1-\frac{(m(t)-\omega'(t))}{2\,\rho}}\,d\omega'd\rho
\ee
Performing the $\rho$ integration first and then using contour integration to carry out the integration over $\omega'$, the imaginary part of the action for an S-wave particle crossing the horizon in an outward direction is found to be
\be 
\Im S = 2\,\pi\,a(t)^2\,\omega\,(5\,m - \frac{\omega}{2})
\ee

\subsection{List of definitions}

\subsubsection{``Planck" and ``shell" language}

We use many words with the prefix ``Planck", most of them are well established, and only very few of which are new; we also use ``shell" several times:

\ha
\underline{Planck} length: $l_{Pl}^{2} \, = \, G_N \hbar /c^{3}$, with $l_{Pl} \, \simeq \, 1.6 \cdot 10^{-33} \, {\rm cm}$.

\ha
\underline{Planck} time: $\tau_{Pl}^2 \, = \, G_N \hbar/c^{5}$, with $\tau_{Pl} \, \simeq \, 5.4 \cdot 10^{-44} \, {\rm s}$.

\ha
\underline{Planck} area: $\sigma_{Pl} \, = \, \pi G_N \hbar /c^{3}$, with $\sigma_{Pl} \, \simeq \, 8.2 \cdot 10^{-66} \, {\rm cm^2}$.

\ha
\underline{Planck} mass:  $m_{Pl}^{2} \, = \, \hbar c /G_N$, with $m_{Pl} \, \simeq \, 2.2 \cdot 10^{-5} \, {\rm g}$.

\ha
\underline{Planck} energy:  $E_{Pl}^{2} \, = \, \hbar c^5 /G_N$, with $m_{Pl} \, \simeq \, 2.0 \cdot 10^{16} \, {\rm erg}$.

\underline{Planck} spectrum:  The standard Planck spectrum for even spin particles.

\ha 
\underline{Planck} sea:  The background field, assumed here to have a Planck spectrum at maximal density and energy.

\underline{Planck} \underline{shell}:  The surface corresponding to a black hole.

\ha
\underline{Shell} front: The surface of the graviton shell flying out from the formation of a black hole, or the merger of two black holes.

\ha
Horizon \underline{shell}:  The surface of interaction with the background at the horizon during inflation.

\subsubsection{Redshift Formulation of Time and Distance}

In order to evaluate the behavior of gravitons we need to use integrals which allow an extrapolation into the future, implying $(1+z) \, -> \, 0$.  We set the low redshift to $z$, and the high redshift to $z_{\star}$,   the redshift at which we assume the main gravitons to be produced.

The expression for the time can be written the normal way
\be
t \; = \; \int_{z}^{z_{\star}} \frac{d z'}{(1+z') \, H(z')}\, ,
\ee
where $1+z$ can have any value above zero.  The time into the future diverges with the natural logarithm of the redshift expression $1+z$ going to zero.  The derivative of redshift $z$ with respect to time can be written as
\be
\frac{d z}{d t} \; = \; - (1+z) H(z)\, .
\ee

The corresponding integral for the luminosity distance between redshift $z$ and redshift $z_{\star}$ can be written as
\be
d_L \; = \; \frac{1+z_{\star}}{1+z} \, \int_{z}^{z_{\star}} \frac{c \, d z'}{ H(z')} \; = \; \frac{1+z_{\star}}{1+z} \, r(z, z_{\star})\, ,
\ee
and again this is valid into the future.  We use redshift $z_{\star}$ as the redshift, when a black hole was formed, or a merger between two black holes happened, so in our interpretation the epoch, when the shell front was initiated in the last few Planck times before what is normally called horizon formation.  Similarly the luminosity distance diverges as $(1+z)^{-1}$ in the future.

\end{document}